\documentclass[useAMS,usenatbib]{mn2e}
\pdfoutput=1
\usepackage{graphicx}
\usepackage{txfonts}
\newcommand{\kms}{\hbox{${\rm km\;s}^{-1}$}}
\newcommand\ion[2]{#1$\;${\small\rmfamily\@Roman{#2}}\relax}%
\def\nai{Na\,{\sc I}}
\def\cai{Ca\,{\sc I}}
\def\fei{$<$Fe\,{\sc I}$>$}
\def\feia{Fe\,{\sc I}\,A}
\def\feib{Fe\,{\sc I}\,B}

\def\hii{H$_2$}

\newcommand{\ha}{H$\alpha$}

\title[The small black holes in NGC\,3368 and NGC\,3489]{Do black hole
masses scale with classical bulge luminosities only? The case of the
two composite pseudobulge galaxies NGC\,3368 and NGC\,3489\thanks{Based on
observations at the European Southern Observatory VLT (078.B-0103(A)),
and on service observations made with the William Herschel Telescope
operated on the island of La Palma by the Isaac Newton Group in the
Observatorio del Roque de los Muchachos of the Instituto de
Astrof\'{i}sica de Canarias.}}  \author[N. Nowak et
al.]{N. Nowak$^{1,2}$\thanks{E-mail: nnowak@mpe.mpg.de},
J. Thomas$^{1,2}$, P. Erwin$^{1,2}$, R. P. Saglia$^{1,2}$,
R. Bender$^{1,2}$, R. I. Davies$^{1}$\\ $^{1}$Max--Planck--Institut
f\"{u}r extraterrestrische Physik, Giessenbachstrasse, 85748 Garching,
Germany\\ $^{2}$Universit\"{a}tssternwarte, Scheinerstrasse 1, 81679
M\"{u}nchen, Germany}

\begin{document}


\pagerange{\pageref{firstpage}--\pageref{lastpage}} \pubyear{2009}

\maketitle

\label{firstpage}

\begin{abstract}
It is now well established that all galaxies with a massive bulge
component harbour a central supermassive black hole (SMBH). The mass
of the SMBH correlates with bulge properties such as the bulge mass
and the velocity dispersion, which implies that the bulge and the
central black hole of a galaxy have grown together during the
formation process. As part of an investigation of the dependence of
the SMBH mass on bulge types and formation mechanisms, we present
measurements of SMBH masses in two pseudobulge galaxies. The spiral
galaxy NGC\,3368 is double-barred and hosts a large pseudobulge with a
tiny classical bulge component at the very centre. The S0 galaxy
NGC\,3489 has only a weak large-scale bar, a small pseudobulge and a
small classical bulge. Both galaxies show weak nuclear activity in the
optical, indicative of the presence of a supermassive black hole.  We
present high resolution, adaptive-optics-assisted, near-infrared
integral field data of these two galaxies, taken with SINFONI at the
Very Large Telescope, and use axisymmetric orbit models to determine
the masses of the SMBHs.  The SMBH mass of NGC\,3368, averaged over
the four quadrants, is $\langle
M_\bullet\rangle=7.5\times10^{6}$~M$_\odot$ with an error of
$1.5\times10^{6}$~M$_\odot$, which mostly comes from the
non-axisymmetry in the data.  For NGC\,3489, a solution without black
hole cannot be excluded when modelling the SINFONI data alone, but can
be clearly ruled out when modelling a combination of SINFONI, OASIS
and SAURON data, for which we obtain
$M_\bullet=(6.00^{+0.56}_{-0.54}|_{\mathrm{stat}}\pm0.64|_{\mathrm{sys}})\times10^{6}$~M$_\odot$. Although
both galaxies seem to be consistent with the $M_\bullet$-$\sigma$
relation, at face value they do not agree with the relation between
bulge magnitude and black hole mass when the total bulge magnitude
(i.e., including both classical bulge and pseudobulge) is considered;
the agreement is better when only the small classical bulge components
are considered. However, taking into account the ageing of the stellar population could change this conclusion.

\end{abstract}


\section{Introduction}

\begin{table*}
 \centering
 \begin{minipage}{140mm}
  \caption{Properties of the galaxies NGC\,3368 and NGC\,3489. As both galaxies host a composite bulge, the effective radius $R_\mathrm{e}$ and the $K$-band bulge magnitude $M_K$ are given for the photometric bulge and the classical bulge component.}\label{tab:parameters}
  \begin{tabular}{cccccccccc}
  \hline
Galaxy & Type & D & PA & $i$ & $R_{e}$ (phot.) & $M_{K}$ (phot.) & $R_{e}$ (class.) & $M_{K}$ (class.) &  Activity\footnote{\citet{Ho-97}}\\
       &      & (Mpc) & ($\degr$) & ($\degr$) & (arcsec) & & (arcsec) & \\
\hline
NGC\,3368 & SAB(rs)ab  & 10.4 & 172 & 53  & 24.9 & -23.42  & 1.6 & -19.48 & L2   \\
NGC\,3489 & SAB(rs)0+  & 12.1 &  71 & 55  &  4.3 & -21.91  & 1.3 & -20.60 & T2/S2   \\
\hline
\end{tabular}
\end{minipage}
\end{table*}

Bulges located in the central regions of disc galaxies are commonly
identified as the region where the excess light above the outer
exponential disc dominates the surface brightness profile. Bulges were
generally regarded as scaled-down versions of elliptical galaxies,
probably formed via minor galaxy mergers. There is now evidence that
there is also a second type of central structure, the so-called
pseudobulges, which are more similar to mini-discs than to
mini-ellipticals. They were first introduced by \citet{Kormendy-93},
and \citet{Kormendy-04b} review properties and formation mechanisms
and present a number of examples. Pseudobulges are thought to be the
result of secular evolution and can be identified e.g. through the
presence of disc-like structure (nuclear spirals, bars or rings),
flattening similar to that of the outer disc, rotation-dominated
kinematics, exponential surface brightness profiles or young stellar
populations. As the formation mechanisms of classical and pseudobulges
are fundamentally different and can happen independently, galaxies
could harbour both types of bulges
\citep{Erwin-03,Athanassoula-05,Erwin-08}. But this fundamental
difference between the formation mechanisms also leads to the question
whether and how a central black hole grows inside a pseudobulge and
how the mass of the black hole relates to pseudobulge
properties. Supermassive black holes (SMBHs) in elliptical galaxies
and classical bulges are known to follow tight correlations with
luminosity (e.g. \citealt{Kormendy-95,MarconiHunt-03}), mass
\citep{Haering-04} and velocity dispersion ($M_\bullet$-$\sigma$
relation, \citealt{Gebhardt-00a,Ferrarese-00}) of the bulge. It is not
clear whether pseudobulges follow the same relations, for several
reasons: 1. There are only very few direct SMBH mass measurements in
pseudobulges. 2. The concept of pseudobulges is relatively new and the
classification criteria therefore differ somewhat from author to
author. 3. The fact that at least some galaxies could contain both
bulge types (composite bulges) makes the classification and
correlation studies even more complicated. The composite bulges need
to be decomposed properly in order to find out with which property of
which bulge component the SMBH mass correlates. \citet{Kormendy-01}
did not find any dependence of the $M_\bullet$-$\sigma$ relation on
the mechanism that feeds the black hole. In contrast \citet{Hu-08}
finds that the black holes in pseudobulges have systematically lower
masses than black holes in classical bulges and ellipticals with the
same velocity dispersion. Both studies suffer from small number
statistics and unclear classification issues.
For low-mass galaxies without classical bulge (i.e. likely hosts
of a pseudobulge) and with virial SMBH mass estimates 
\citet*{Greene-08} found no deviation from the $M_\bullet$-$\sigma$ relation,
but a likely disagreement with the $M_\bullet$-$M_{\mathrm{bulge}}$
relation. \citet{Gadotti-08} found for a large number of
SDSS galaxies that pseudobulges, classical bulges and ellipticals
cannot follow both the $M_{\bullet}$-$\sigma$ \emph{and} the
$M_{\bullet}$-$M_{\mathrm{bulge}}$ relation at the same time. As they
estimated $M_{\bullet}$ from the $M_{\bullet}$-$M_{\mathrm{bulge}}$
relation by \citet{Haering-04} it is not clear whether their
pseudobulges follow a different $M_{\bullet}$-$\sigma$ or
$M_{\bullet}$-$M_{\mathrm{bulge}}$ relation or both.

In this paper we present a thorough analysis and derivation of the
black-hole masses via extensive stellar dynamical modelling of
NGC\,3368, a double-barred spiral galaxy \citep{Erwin-04} of type
SAB(rs)ab with a well-defined pseudobulge and a very small classical
bulge component, and NGC\,3489, a SAB(rs)0+ galaxy with a weak large
bar, a small pseudobulge and a similar-sized classical bulge. All
important parameters of the two galaxies are listed in Table
\ref{tab:parameters}. Using high-resolution imaging we are able to
identify and decompose the pseudobulge and classical bulge
components. High-resolution adaptive-optics assisted near-IR
integral-field spectroscopy enables us to model each quadrant
separately. In contrast to our two previous studies of elliptical
galaxies \citep{Nowak-07,Nowak-08}, non-axisymmetries may play a
larger role due to the barred nature of the galaxies.
 
The nucleus of NGC\,3368 is weakly active and can be classified as a
LINER2 based on optical emission line ratios
\citep*{Ho-97}. \citet{Maoz-05} and \citet{Maoz-07} report long-term UV
variations, which suggests the presence of an AGN and thus a
SMBH. NGC\,3489 has a weak LINER/HII transition type or Seyfert 2
nucleus \citep{Ho-97}.

We adopt a distance to NGC\,3368 of $10.4$~Mpc throughout the paper
based on surface brightness fluctuation measurements
\citep{Tonry-01}. At this distance, 1~arcsec corresponds to
$\sim50$~pc. For NGC\,3489 we adopt a distance of 12.1~Mpc
($\sim59$~pc~arcsec$^{-1}$), also based on the measurements of
\citet{Tonry-01}. If the $M_\bullet$-$\sigma$ relation
\citep{Tremaine-02} applies, both galaxies are close enough to resolve
the sphere of influence of the black hole from the ground with
adaptive optics.

This paper is organised as follows: In \S 2 we discuss the morphology
of the two galaxies, including photometric (and spectroscopic)
evidence for pseudobulges and classical bulges. The spectroscopic
data, including stellar kinematics, gas kinematics and line strength
indices are described in \S 3. The stellar dynamical modelling
procedure and the results for the SMBH mass of each galaxy are
presented in \S 4 and \S 5. \S 6 summarises and discusses the
results.

\section{Imaging}

\subsection{Identifying Discs, Pseudobulges, and Classical Bulges}

\begin{figure*}
\centering
\includegraphics[width=0.7\linewidth,keepaspectratio]{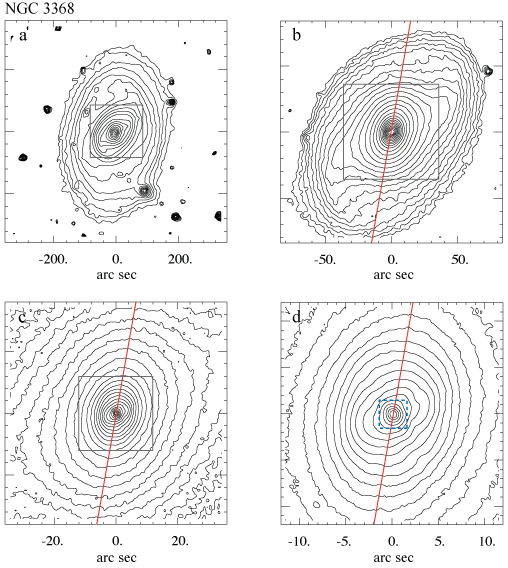}
\caption{Isophotal maps of NGC\,3368 on different scales, all with
logarithmic intensity scaling.  (a) SDSS $r$-band isophotes, showing
the outer disc.  The gray box outlines the region shown in the next
panel. (b) $K$-band isophotes from the image of \citet{Knapen-03},
showing the outer bar and the lens just outside.  (c) $K$-band
isophotes, showing the region inside the outer bar, including the
bright inner pseudobulge region.  (d) $K$-band isophotes, showing the
elliptical contours of the inner pseudobulge, the inner bar, and the
central bulge inside.  The dashed blue square shows the approximate
field of view of our 100mas SINFONI observations.  For reference, we
indicate the adopted major-axis position angle ($172\degr$) with the
diagonal red line in panels b--d.}
\label{fig:isophotes-n3368}
\end{figure*}

As mentioned in the Introduction, both NGC\,3368 and NGC\,3489 have
complex morphologies; they are not simply an exponential disc plus a
classical bulge. In this section, we explain our approach for analysing
the morphology of these galaxies, and how we decompose their surface
brightness profiles for purposes of dynamical modelling.

Since discs and bulges can have different mass-to-light ratios
$\Upsilon$, we need to separate these components for modelling
purposes.  But we also need to ensure that what we call the ``bulge''
really is distinct from the disc, and not simply a
higher-surface-brightness extension of the disc.  NGC\,3368 has already
been classified by \citet{Drory-07} as a galaxy where the bulge is a
``pseudobulge'', so we need to consider whether this galaxy even has a
distinct bulge; if it does have a pseudobulge, we need to identify
\textit{that}.

\begin{figure*}
\includegraphics[width=0.8\linewidth,keepaspectratio]{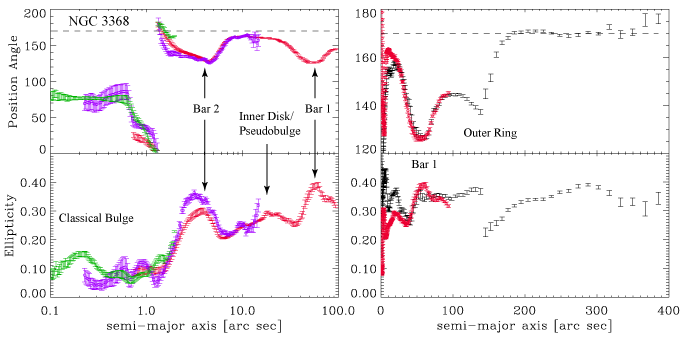}
\caption{Isophotal ellipse fits for NGC\,3368, plotted on logarithmic
(left panel) and linear (right panel) scales. Black points are for the
SDSS $r$-band image, red points are for the $K$-band image of
\citet{Knapen-03}, purple points are for the NICMOS2 F160W image and
green points are for the SINFONI image (derived from our datacube).
Major morphological features are marked (``Bar 1'' = outer bar, ``Bar
2'' = inner bar).  The dashed line in the top panels indicates our
adopted position angle ($172\degr$) for the galaxy disc.}
\label{fig:efits-n3368}
\end{figure*}

Our overall approach is as follows: First, we perform a ``naive''
global bulge/disc decomposition, where we treat the entire
surface-brightness profile as the combination of an outer exponential
(the ``main disc'') and a S\'ersic profile. The latter component is
the ``photometric bulge,'' which would traditionally be considered
\textit{the} bulge -- i.e., the kinematically hot spheroid -- of the
galaxy. Second, we focus on the photometric bulge region (that is,
where the S\'ersic component dominates the modelled global light
profile), and examine the morphology and stellar kinematics. In both
galaxies, we find evidence that these regions are predominantly
disclike and thus \textit{not} classical bulges. This argues that we
should treat most of the photometric bulge region as part of the
disc. Finally, we also find evidence that the central few hundred
parsecs contain an additional component: a central light excess above
the disclike bulge region associated with kinematically hot stellar
kinematics and rounder isophotes.  It is this last component which we
call the ``classical bulge,'' and which we treat as a separate stellar
component in our modelling.

We thus consider both galaxies to be similar to NGC 2787 and NGC 3945,
where \citet{Erwin-03} showed that the photometrically defined
``bulges'' of both galaxies were really bright ``inner discs,'' with
much smaller (and rounder) bulges \textit{inside} the inner discs. 
\citet{Erwin-08} and Erwin et al.\ (2009, in prep) discuss more examples
of such ``composite bulge'' galaxies, and \citet{Athanassoula-05}
provides a theoretical context for such systems.

To summarise: Both galaxies appear to consist of a disc with an outer
exponential profile\footnote{This is a simplification, since the profile
of the disc at very \textit{large} radii may change
\citep{Erwin-08b}, but this has no effect on our modelling.}
and a steeper inner profile, along with a small central excess with
rounder isophotes and stellar kinematics which we consider to be the
(classical) bulge.  The steep inner part of the disc can be considered a
discy pseudobulge, but for modelling purposes we treat it as just the
inner part of the disc.

\subsection{Imaging Data and Calibrations} 

The imaging data we use comes from a variety of sources, including the
Two-Micron All-Sky Survey \citep[2MASS;][]{Skrutskie-06}, the Sloan
Digital Sky Survey \citep[SDSS;][]{York-00}, and the \textit{Hubble
Space Telescope} archive.  We also use near-IR images taken with the
Isaac Newton Group Red Imaging Device (INGRID, a $1024^{2}$ near-IR
imager with 0.24~arcsec pixels) on the William Herschel Telescope: a
$K$-band image of NGC\,3368 from \citet{Knapen-03}, available via NED,
and an $H$-band image of NGC\,3489 obtained during service/queue time
(February 11, 2003).  The seeing for the INGRID images was 0.77~arcsec
FWHM for NGC\,3368 and 0.74~arcsec FWHM for NGC\,3489.  Finally, we
also use $K$-band images created from our SINFONI datacubes (see
Fig. \ref{fig:image}).

The SDSS $r$-band images are used for measuring the shape of the
outer-disc isophotes, which helps us determine the most likely
inclination for each galaxy.  The innermost isophote shapes and
surface brightness profiles are determined from the \emph{HST}
archival images.  For NGC\,3368, we choose a NICMOS2 F160W image (PI
Mulchaey, proposal ID 7330) for that purpose; with F450W and
F814W WFPC2 images (PI Smartt, proposal ID 9042) we construct
high-resolution colour maps and attempt to correct the NICMOS2 image
for dust extinction.  For NGC\,3489, we used F555W and F814W WFPC2
images (PI Phillips, proposal ID 5999).  We attempted to correct the
NICMOS2 F160W image (for NGC\,3368) and the WFPC2 F814W image (for
NGC\,3489) for dust extinction, following the approach of
\citet{Barth-01} and \citet{Carollo-97}.  This involved creating a $V
- H$ colourmap for NGC\,3368 and a $V - I$ colourmap for NGC\,3489,
then generating corresponding $A_H$ and $A_I$ extinction maps and
correcting the NICMOS2 and WFPC2 F814W images.  The results were
reasonably successful for NGC\,3489, but less so for NGC\,3368,
perhaps due to the much stronger extinction in the latter galaxy.

The 2MASS images are used primarily to calibrate the INGRID near-IR
images. Since the latter suffer from residual sky-subtraction problems,
we calibrate them by matching surface-brightness profiles from the
INGRID images with profiles from the appropriate 2MASS images ($K$-band
for NGC\,3368, $H$-band for NGC\,3489), varying both the scaling and a
constant background offset until the differences between the two
profiles are minimized.  We then carry over this calibration to
surface-brightness profiles from the \emph{HST} images: i.e., we calibrate the
NICMOS2 F160W profile to $K$-band for NGC\,3368 by matching it to the
(calibrated) INGRID $K$-band profile (including a background offset),
and similarly match the WFPC2 F814W profile to INGRID $H$-band profile
for NGC\,3489.  Profiles from the SINFONI $K$-band images are then
calibrated by matching them to the appropriate calibrated \emph{HST} profiles.

\subsection{NGC\,3368}

\subsubsection{Morphological overview}

NGC\,3368 is a relatively complex spiral galaxy, with a number of
different stellar components.  \citet{Erwin-04} argued that the central
regions of NGC\,3368 included at least three distinct components: an
outer bar with semi-major axis $a\approx61$--$75$~arcsec (4.4--5.4~kpc,
deprojected), an ``inner disc'' extending to $a\approx21$--$30$~arcsec
($1.1$--$1.6$~kpc, deprojected), and an inner bar with
$a\approx3.4$--$5.0$~arcsec (200--300~pc, deprojected).  As noted above,
this set of nested structures is very similar to that of the
double-barred galaxy NGC\,3945 \citep{Erwin-99}, where \citet{Erwin-03}
found that the galaxy's ``photometric'' bulge could be decomposed into a
bright, kinematically cool disc (first noted by \citealt{Kormendy-82})
with an exponential profile and a much smaller, rounder object
dominating the inner few hundred parsecs -- apparently a central,
spheroidal bulge.

The isophotes of NGC\,3368 are shown in
Fig. \ref{fig:isophotes-n3368} for different scales.  The isophotal
ellipse fits to ground-based and \emph{HST} near-IR images are shown
in Fig. \ref{fig:efits-n3368}.  In the inner region
($r\sim3-4$~arcsec), the ellipticity rises to a local maximum of
$\sim0.3$ and the isophotes are closely aligned with the outer
disc. Even further in, inside the inner bar (semi-major axis
$a<2$~arcsec), the isophotes become quite round, with a mean
ellipticity of $\approx 0.1$. The isophotes in this region also twist
significantly; inspection of both the NICMOS2 image and our SINFONI
datacubes indicate that this twisting is produced by strong dust lanes
on either side of the galaxy centre.  The true (unextincted)
ellipticity in this region is probably close to 0. This suggest that
NGC\,3368 harbours a small classical bulge, again in analogy to
NGC\,3945.  The size of the classical bulge region is much larger than
the NICMOS PSF and is therefore well resolved and an AGN can thus be
excluded. In addition we do not find emission lines characteristic for
an AGN in our SINFONI spectra (see below).


\subsubsection{Bulge-disc decomposition}

\begin{figure}
\includegraphics[width=0.95\linewidth,keepaspectratio]{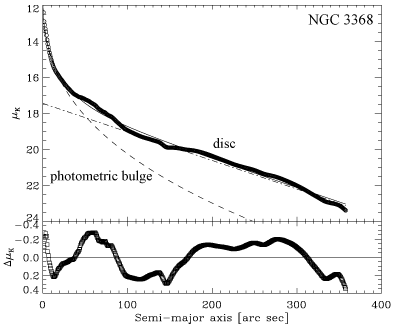}
\caption{Global bulge-disc decomposition of NGC\,3368. The data points
(circles) are the major-axis $K$-band profile, combining the
\emph{HST}-NICMOS2 F160W image ($r<7.1$~arcsec) with the $K$-band image of
\citet{Knapen-03} for $7.2<r<86$~arcsec and the SDSS $r$-band
image for $r>86$~arcsec, all calibrated to $K$. Also shown is the
best S\'ersic + exponential fit to the data and the residuals (bottom
panel).  The S\'ersic component represents the ``photometric bulge,''
which dominates the light at $r<50$~arcsec. (Most of this, we argue, is
really a part of the disc; see text for details.)}
\label{fig:global-decomp-n3368}
\end{figure}

\begin{figure}
\includegraphics[width=0.95\linewidth,keepaspectratio]{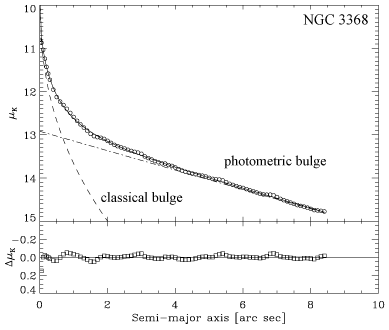}
\caption{Bulge-disc decomposition of the inner photometric bulge of
NGC\,3368.  The data points (circles) are the major-axis $K$-band
profile for the inner $r<8.5$~arcsec, combining our SINFONI data cube
($r<0.3$~arcsec) with the \emph{HST}-NICMOS2 F160W image, both calibrated to
$K$.  Also shown is the best S\'ersic + exponential fit to the data at
$r \leq 8$~arcsec and the residuals (bottom panel); in this fit, the
S\'ersic component represents the ``classical bulge,'', while the
exponential is the inner part of the photometric bulge -- that is, it is
the steep inner part of the galaxy disc.}
\label{fig:inner-decomp-n3368}
\end{figure}

Fig.~\ref{fig:global-decomp-n3368} shows a global
bulge-disc decomposition for NGC\,3368, in which the photometric bulge
(the S\'ersic component of the fit) dominates the light at
$r<50$~arcsec.

Fig.~\ref{fig:inner-decomp-n3368} shows a S\'ersic + exponential
decomposition of the inner $r<10$~arcsec region. In this
decomposition, we are now treating what we previously identified as
the photometric bulge as a disc-like component (i.e. the
pseudobulge), which has (compared to the outer disc) a relatively
steep exponential profile, plus a central S\'ersic excess. Note that
we are fitting the inner region of the original data, i.e. we did not
subtract the outer exponential disc. The result is a reasonably
good fit, suggesting that the inner $r<2$~arcsec region -- where, as
noted, the isophotes are quite round -- is a separate component
(best-fit S\'ersic parameters: $n = 2.35$, $R_e = 1.60$~arcsec, $\mu_e
= 14.53$).

\subsubsection{Kinematic structure of the photometric bulge}

\begin{figure}
\includegraphics[width=0.9\linewidth,keepaspectratio]{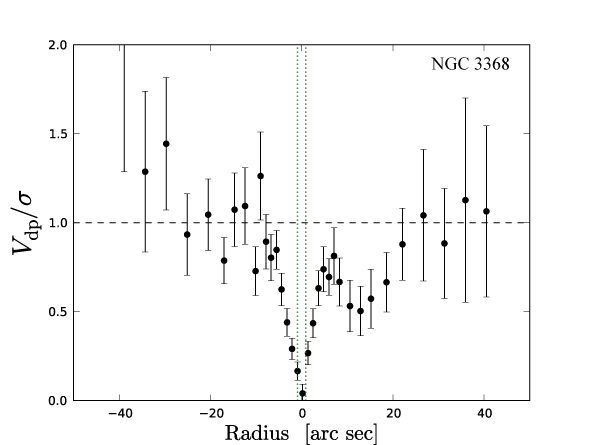}
\caption{Local estimates of the ratio of ordered motion to random
motion in stellar kinematics along the major axis of NGC\,3368, within
the photometric bulge region ($r < 50$~arcsec), based on the long-slit
data of \citet{Heraudeau-99}.  We first deproject the stellar
velocities to their in-plane values (correcting from the observed PA
of $5\degr$ to our adopted major-axis PA of $172\degr$), then divide
them by the observed velocity dispersion values.  Ratios $< 1$
indicate a dominance of velocity dispersion over bulk rotation; ratios
$> 1$ indicate kinematically cooler regions. The vertical green
dotted lines indicate the seeing of the original spectroscopic
observations (FWHM $= 1.8$~arcsec).}
\label{fig:v-div-sigma-n3368}
\end{figure}

In Fig.~\ref{fig:v-div-sigma-n3368} we use the long-slit kinematic
data of \citet{Heraudeau-99} to show an estimate of the local ratio of
ordered to random stellar motions as a function of radius: $V_{\rm
dp}$, which is the observed stellar velocity deprojected to its
in-plane value (assuming an axisymmetric velocity field), divided by
the observed velocity dispersion at the same radius.  Even though the
data are all inside the photometric bulge ($r < 50$~arcsec), the ratio
of $V_{\rm dp}/\sigma$ rises above 1 over much of this region.  This
is certainly higher than one would expect for a classical
(kinematically hot) bulge, in which stellar motions are dominated by
velocity dispersion.  An unpublished spectrum with higher S/N from the
Hobby-Eberly Telescope (M.  Fabricius, private communication) shows
even larger values of $V_{\rm dp}/\sigma$ for $r > 30$~arcsec, as well
as $V_{\rm dp}/\sigma > 1$ on both sides of the centre at $r \sim
5$--9~arcsec.  Our tentative conclusion is that most of the
photometric bulge is thus a discy, kinematically cool pseudobulge (in
effect, an inward extension of the disc), similar to that found in
NGC\,3945. We note that the photometric bulge of NGC\,3368 has also
been classified as a pseudobulge by \citet{Drory-07}, based on
morphological features in \emph{HST} images. Dynamical modelling of
the high-resolution SINFONI data shows that the centre of the
photometric bulge harbours a kinematically hot component (see \S 4.4),
thus confirming the presence of a small classical bulge component.

\begin{figure*}
\includegraphics[width=1.0\linewidth]{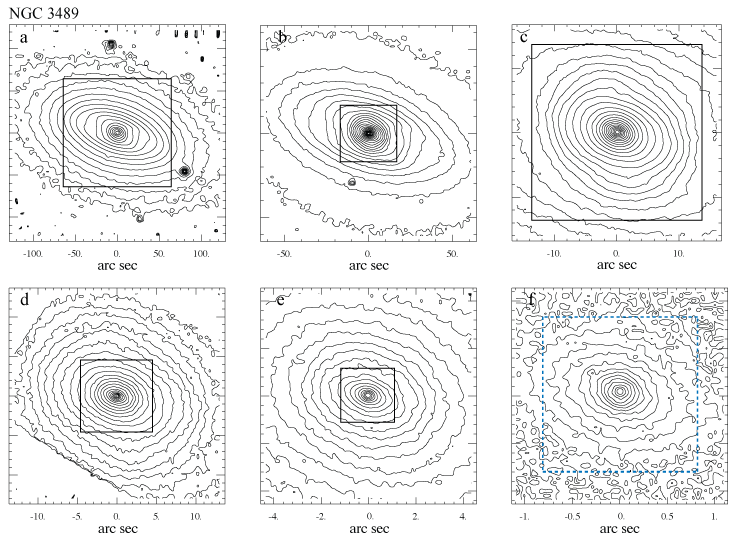}
\caption{Isophotal maps of NGC\,3489 on different scales, all with
logarithmic intensity scaling.  Each successive panel is a zoom of the
region outlined with a box in the preceding panel. (a) SDSS
$r$-band isophotes, showing the outer disc.   (b) $H$-band
isophotes from our WHT-INGRID image.  (c) Same, now showing the
bar oriented almost vertically. (d) Now showing dust-corrected
F814W isophotes from the PC chip of the WFPC2 image.  (e) Same
as previous, showing the inner part of the pseudobulge region.
(f) Same as previous, but now showing the classical bulge region
and the possible nuclear disc.  The dashed blue square shows the
approximate field of view of our 25mas SINFONI observations.}
\label{fig:isophotes-n3489}
\end{figure*}

\subsubsection{Orientation and inclination of the galaxy}

Our ellipse fits of the merged SDSS $r$-band image (black points in
Fig.~\ref{fig:efits-n3368}) shows a consistent position angle of
$\approx 172\degr$ for the outermost isophotes.  These ellipse fits
extend well outside the star-forming outer ring ($r\sim180$~arcsec) and
are thus unlikely to be affected by any intrinsic noncircularity of
the ring itself.  This position angle agrees very well with the
\textit{kinematic} position angles determined from both H\,\textsc{i}
observations (PA~$\approx170\degr$, based on the data of
\citealt{Schneider-89}, as reported by \citealt{Sakamoto-99}) and from
the Fabry-Perot \ha+[N\,\textsc{ii}] velocity field of
\citet[][see also \citealt*{Moiseev-04}]{Silchenko-03}.  Sil'chenko et
al.\ also find a kinematic position angle of 170--$175\degr$
in the stellar kinematics of the inner $2-5$~arcsec, from their IFU
data.

The ellipticity of the outer $r$-band isophotes is $\approx0.37$,
with a range of 0.34--0.39.  A lower limit on the inclination is thus
$51\degr$, for a razor-thin disc; thicker discs imply higher
inclinations. For an intrinsic thickness of $c/a = 0.2$--0.25, the
inclination is $i \approx 53\degr$.   This is close to the
inclination of $50\degr$ estimated by \citet*{Barbera-04}, based on
Fourier analysis of \citet{Frei-96} images (note that these images do
not extend beyond the outer-ring region, and so they might in
principle be biased if the ring is noncircular). 

An additional, independent estimate of the inclination can be had by inverting
the Tully-Fisher relation: since we know the observed H\,\textsc{i} velocity width
\textit{and} the distance to NGC\,3368, we can determine the inclination needed
to make the galaxy follow the Tully-Fisher relation.  We use the recently
published 2MASS Tully-Fisher relation of \citet*{Masters-08}.  For a 2MASS
$K$-band ``total'' magnitude \citep{Jarrett-03} of 6.31 (including a slight
reddening correction from \citealt*{Schlegel-98}, as given
by NED) and a distance of 10.4 Mpc, the absolute magnitude is $M_K = -23.795$.
Using the ``Sb'' T-F relation from table~3 of \citet{Masters-08}, this corresponds
to a corrected, edge-on velocity width of $W_{\rm corr} = 425$ \kms.  For the
observed width, we use the tabulated value in \citet{Springob-05}, which is
$W_c = 324$ \kms; after applying the recommended correction for turbulent
broadening (6.5 \kms), this becomes $W_{\rm corr} \sin i = 317.5$ \kms, and
thus $i = 48\degr$.

Taken all this into considerations, we can argue that NGC\,3368 has a
line of nodes with PA $\approx 172\degr$ and an inclination somewhere
between $48\degr$ and $55\degr$, most likely $\approx 53\degr$.

\begin{figure*}
\includegraphics[width=0.8\linewidth]{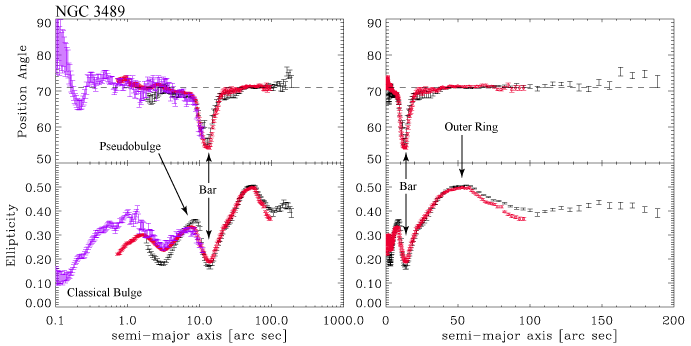}
\caption{Isophotal ellipse fits for NGC\,3489, plotted on logarithmic
(left panel) and linear (right panel) scales. Black points are for the
SDSS $r$-band image, while red points are for the INGRID $H$-band
image and purple points are for the dust-corrected WFPC2 F814W
image. Major morphological features are indicated. Note that due to
projection effects, the bar shows up as both a strong twist in the
position angle and a \textit{minimum} in the ellipticity. The dashed
line in the top panels indicates our adopted position angle
($71\degr$) for the galaxy disc.}
\label{fig:efits-n3489}
\end{figure*}

\subsection{NGC\,3489}

\subsubsection{Morphological overview and evidence for a composite pseudobulge} 

NGC\,3489 is structurally somewhat simpler than NGC\,3368, with only one
bar instead of two.  The bar itself is rather weak and difficult to
recognise, because it lies almost along the minor axis of the galaxy.
Projection effects thus foreshorten it so that it is visible primarily
due to the abrupt isophote twists, manifesting in the ellipse fits as an
extremum in the fitted position angle and a \textit{minimum} in the
ellipticity (Fig.~\ref{fig:isophotes-n3489} and
Fig.~\ref{fig:efits-n3489}). Further outside, the isophotes become
maximally elongated at $r \sim 50$~arcsec, and then converge to a mean
ellipticity of $\approx 0.41$ at larger radii.  As shown by
\citet{Erwin-03b}, the ellipticity peak at $r \sim 50$~arcsec is due to an
outer ring; the lower ellipticity outside is thus the best
representation of the outer disc.

\subsubsection{Bulge-disc decomposition}

\begin{figure}
\includegraphics[width=1.0\linewidth,keepaspectratio]{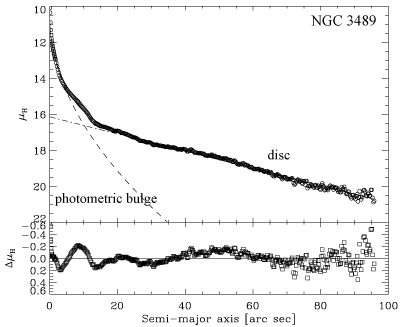}
\caption{Global bulge-disc decomposition of NGC\,3489. The data points
(circles) are the major-axis cut, combining our ground-based $H$-band
image ($r > 1.7$~arcsec) with the dust-corrected WFPC2 F814W image
(scaled to $H$-band). Also shown is the best S\'ersic + exponential
fit to the data and the residuals (bottom panel). The S\'ersic
component represents the ``photometric bulge,'' which dominates the
light at $r < 10$~arcsec. (As with NGC~3368, we argue that most of this
is really part of the disc; see text for details.)
\label{fig:global-decomp-n3489}}
\end{figure}

\begin{figure}
\includegraphics[width=1.0\linewidth,keepaspectratio]{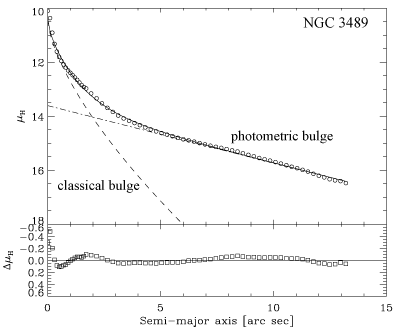}
\caption{Bulge-disc decomposition of the inner photometric bulge of
NGC\,3489.  The data points (circles) are the $H$-band profile from
Fig.~\ref{fig:global-decomp-n3489} for the inner $r < 13$~arcsec (with
data at $r < 1.7$~arcsec coming from the dust-corrected WFPC2 F814W
image). Also shown is the best S\'ersic + exponential fit to the data
at $r \leq 13$~arcsec and the residuals (bottom panel), with the
S\'ersic component representing the classical bulge.}
\label{fig:inner-decomp-n3489}
\end{figure}

We do find some evidence that the inner structure of NGC~3489 is similar
to that of NGC~3368 (except for the absence of an inner bar in
NGC~3489). We start with the global bulge-disc decomposition
(Fig.~\ref{fig:global-decomp-n3489}), which results in a S\'{e}rsic
component (the photometric bulge) dominating the light at $r \leq
10$~arcsec. The isophotes in this region are still fairly elliptical --
e.g., ellipticity $\approx 0.33$ at $r \sim 6$--8~arcsec, which suggests
that the photometric bulge is a flattened structure.

As in the case of NGC\,3368, we find that the profile of the inner
photometric bulge can be decomposed into an exponential plus a
smaller, additional S\'{e}rsic component
(Fig.~\ref{fig:inner-decomp-n3489}).  This S\'{e}rsic component
dominates the light at $r \leq 2$~arcsec.  Note, however, that the
isophotes do become quite elliptical at $ r \sim 1$~arcsec, so we do
not have as clean a case as in NGC\,3368 for a rounder spheroidal
component. The fit in Fig.~\ref{fig:inner-decomp-n3489} shows
evidence for a possible nuclear excess at $r < 0.5$ arcsec.  This
might be evidence for a separate nuclear star cluster, similar to that
seen in the profile of NGC~2787 \citep{Erwin-03}.

\subsubsection{Kinematic structure of the photometric bulge}

\begin{figure}
\includegraphics[width=0.9\linewidth,keepaspectratio]{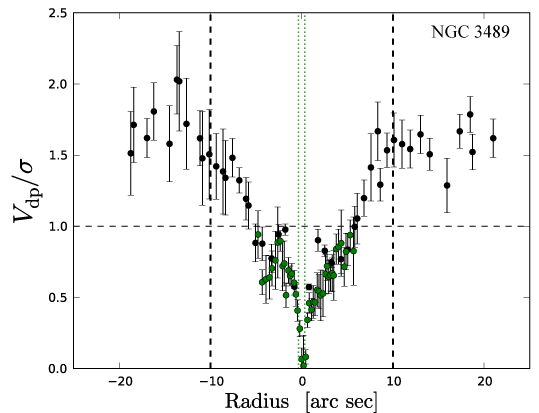}
\caption{Local estimates of the ratio of ordered motion to random motion
in stellar kinematics along the major axis of NGC\,3489, using
major-axis profiles extracted from the velocity and velocity dispersion
fields of the SAURON (black) and OASIS (green) observations; the
vertical green dotted lines indicate the seeing of the OASIS
observations (FWHM $= 0.69$~arcsec).  Stellar velocities are deprojected
to their in-plane values, then divided by the observed velocity
dispersion values.  Ratios $< 1$ indicate a dominance of velocity
dispersion over bulk rotation. At $r \leq 10$~arcsec (vertical dashed
line), the photometric bulge dominates the light (see
Fig.~\ref{fig:global-decomp-n3489}).  Since $V_{\rm dp}/\sigma > 1$
within this region, the outer part of the photometric bulge is still
dominated by rotation, making it a kinematic pseudobulge.
\label{fig:v-div-sigma-n3489}}
\end{figure}

There is in addition kinematic evidence that the exponential part
of the photometric bulge region is kinematically cool and thus a
pseudobulge. In Fig.~\ref{fig:v-div-sigma-n3489} we plot the local
ratio of (deprojected) stellar rotation velocity to velocity
dispersion.  These values are based on synthesised long-slit profiles
derived from the SAURON and OASIS velocity and velocity
dispersion fields \citep{Emsellem-04,McDermid-06}, using slits at PA
$= 71\degr$.  The $V_{\rm dp}/\sigma$ ratio rises to values $> 1$ at
$r \geq 7$~arcsec, still within the photometric bulge-dominated
region, which suggests that the photometric bulge is at least partly
dominated by rotation. As in NGC\,3368 $V_{\rm dp}/\sigma$ drops
in the centre but only at smaller radii. Together with the significant
flattening in the centre this suggests that the very central region of
NGC\,3489 is structurally different (dynamically colder), which is
also supported by the dynamical analysis (see \S 5.5).

\subsubsection{Orientation and inclination of the galaxy} 

Lacking the extensive large-scale gas kinematic information that was
available for NGC\,3368, we rely on the isophotes of the outer disc to
determine the global orientation of NGC\,3489.  Fortunately, apart from
the local maximum in ellipticity at $r \sim 50$~arcsec due to the outer
ring (see above), the outer disc is fairly well defined, with position
angle $= 71\degr$ and a mean ellipticity of 0.41, corresponding to an
inclination of $55\degr$.

\section{Spectroscopy}

\subsection{Data \& Data Reduction}

NGC\,3368 and NGC\,3489 were observed between March 22 and 24, 2007,
as part of guaranteed time observations with SINFONI
\citep{Eisenhauer-sinfoni,Bonnet-sinfoni}, an adaptive-optics assisted
integral-field spectrograph at the VLT UT4.  We used the $K$-band
grating and the $3\times3$~arcsec$^2$ field of view
($0.05\times0.1$~arcsec$^2$~spaxel$^{-1}$) for NGC\,3368 and the
$0.8\times0.8$~arcsec$^2$ field of view
($0.025\times0.0125$~arcsec$^2$~spaxel$^{-1}$) for NGC\,3489. The
total on-source exposure time was $80$~min for NGC\,3368 and
$120$~min for NGC\,3489, consisting of $10$~min exposures taken in
series of ``object--sky--object'' cycles, dithered by a few spaxels.

\begin{figure*}
\includegraphics[width=0.9\linewidth,keepaspectratio]{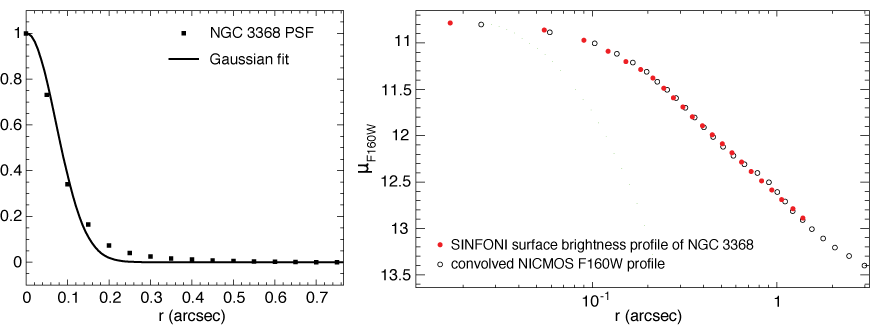}
\caption{Left panel: SINFONI PSF derived by observing a star of the
same magnitude and colour as the nucleus of NGC\,3368 (red dashed
line). A Gaussian fit is overplotted for comparison (black solid
line). Its FWHM is $\sim0.165$~arcsec. Right: Comparison of the
SINFONI $K$-band surface brightness profile with the surface
brightness profile of an \emph{HST} NICMOS2 F160W image convolved with
a Gaussian such that the spatial resolution is $0.165\arcsec$. The
SINFONI profile is shifted such that it matches the NICMOS
profile. \label{fig:psf}}
\end{figure*}

The laser guide star (LGS) PARSEC \citep{Rabien-04,Bonaccini-02} was
used for the AO correction of NGC\,3368, with the tip-tilt sensor
closed on the nucleus of NGC\,3368 ($R=13.58$, $B-R=1.86$ within a
$3$~arcsec diameter aperture). Although the nucleus itself is just
bright enough to be used as natural guide star, its shape is rather
irregular and not pointlike in the $R$-band due to the large amounts
of dust in the nuclear regions and the lack of a strong AGN. Therefore
a better AO correction was expected from using the LGS instead. The
ambient conditions were good and stable, with an average seeing of
$\approx0.6$~arcsec in the near-IR. The point-spread function was
derived by taking an exposure of a nearby star with approximately the
same $R$-band magnitude and $B-R$ colour as the nucleus of NGC\,3368,
using the LGS with the PSF star itself as tip-tilt reference star.
The FWHM of the PSF is $\approx0.165$~arcsec (see left panel of
Fig. \ref{fig:psf}) and the achieved Strehl ratio is
$\approx14\%$. Due to the time gap between the observations of the
galaxy and the PSF star the measured PSF shape could be different from
the PSF during the galaxy observations. We compared the surface
brightness profile of the SINFONI data with the surface brightness
profile of an \emph{HST} NICMOS2 F160W image, convolved with Gaussians
of different widths and found that the NICMOS surface brightness
profile most closely resembles the SINFONI profile for a
$\mathrm{FWHM}\approx0.165$ (see right panel of Fig. \ref{fig:psf}),
confirming our PSF measurement. Note that the Gaussian fitted to
the PSF in the left panel of Fig. \ref{fig:psf} was only used to
determine a nominal spatial resolution and as a reference for
comparison with the NICMOS surface brightness profile. A single
Gaussian does not fit the wings of the PSF, however, the discrepency
between the PSF and the fit is only $\sim3$\% in integrated flux. For
the dynamical modelling we do not use this fit, but the observed image
of the PSF star.

\begin{figure}
\centering
\includegraphics[width=.7\linewidth,keepaspectratio]{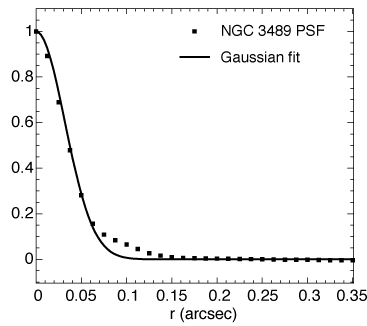}
\caption{PSF of NGC\,3489 with a FWHM of
$\sim0.08$~arcsec. As in Fig. \ref{fig:psf} a Gaussian fit is overplotted for comparison. \label{fig:psf3489}}
\end{figure}

NGC\,3489 was observed using its nucleus with $R=13.22$ ($3$~arcsec
diameter aperture) as natural guide star for the AO correction. A PSF
star with a similar magnitude and $B-R$ colour was observed regularly
in order to determine the spatial resolution. The ambient conditions
were excellent and stable with a seeing around $0.5$~arcsec in the
near-IR, resulting in a FWHM of the PSF of $\approx0.08$~arcsec and a
Strehl ratio of $43\%$ (see Fig. \ref{fig:psf3489}).

\begin{figure}
\includegraphics[width=.9\linewidth,keepaspectratio]{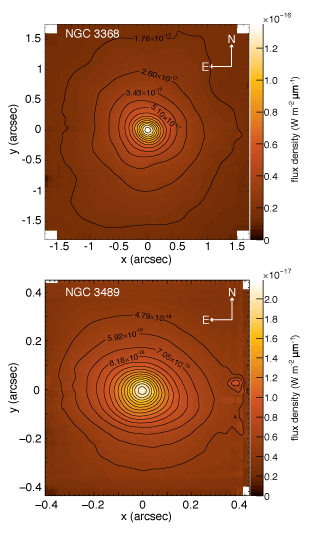}
\caption{SINFONI images with overplotted isophotes of NGC\,3368
(upper panel) and NGC\,3489 (lower panel). The contour levels increase
linearly. \label{fig:image}}
\end{figure}

The data reduction was done using the SINFONI data reduction package
{\sc SPRED} \citep{Schreiber-spred,Abuter-spred} as explained in
\citet{Nowak-08}. The reduction of the telluric standard and the PSF
reference star was done with the ESO pipeline. For the flux
calibration we used the telluric standard stars Hip\,046438
and Hip\,085393 with 2MASS \emph{Ks} magnitudes of $7.373$ and $6.175$
respectively as a reference. Fig. \ref{fig:image} shows the
flux-calibrated images of the two galaxies, collapsed along the
wavelength direction.

\subsection{Stellar Kinematics in NGC\,3368}

The SINFONI data of NGC\,3368 were binned using a binning scheme with
five angular and ten radial bins per quadrant, adopting a major-axis
position angle of $172\degr$. As in \citet{Nowak-08} we used the
maximum penalised likelihood (MPL) technique of \citet{Gebhardt-00c}
to extract the stellar kinematics from the first two CO bandheads
$^{12}$CO(2--0) and $^{12}$CO(3--1), i.e. the spectral range between
$2.279~\umu$m and $2.340~\umu$m rest frame wavelength. With the MPL
method, non-parametric line-of-sight velocity distributions (LOSVDs)
are obtained by convolving an initial binned LOSVD with a linear
combination of template spectra. The residual differences between the
resulting model spectrum and the observed galaxy spectrum are then
calculated. Then the velocity profile and the template weights are
successively adjusted in order to optimise the fit by minimizing the
function $\chi^2_\mathrm{P}=\chi^2+\alpha\mathcal{P}$, where $\alpha$
is the smoothing parameter and $\mathcal{P}$ is the penalty
function. The S/N of the binned spectra ranges between $80$ and $120$
with a mean value of $\sim110$. In order to determine the optimal
smoothing parameters we performed simulations in the same way as in
Appendix B of \citet{Nowak-08}, but tailored to our dataset. For a
galaxy with a velocity dispersion around $100$~\kms, a velocity bin
width of $\sim35$~\kms\ and the mentioned S/N a smoothing parameter
$\alpha\approx5$ is appropriate. As kinematic template stars we chose
four K and M giants which have about the same intrinsic CO equivalent
width (EW) as the galaxy (12--14~{\AA}, using the EW definition and
velocity dispersion correction from \citealt{Silge-03}). The
uncertainties on the LOSVDs are estimated using Monte Carlo
simulations \citep{Gebhardt-00c}. First, a reference galaxy spectrum
is created by convolving the template spectrum with the measured
LOSVD. Then 100 realisations of that initial galaxy spectrum are
created by adding appropriate Gaussian noise. The LOSVDs of each
realisation are determined and used to specify the confidence
intervals. We verified that the error bars are correct by checking
that the S/N in the simulated spectra corresponds to the S/N
measured from the galaxy spectrum. As shown in \citet{Nowak-08},
possible biases in the measured LOSVDs are always smaller than the
statistical errors.

\begin{figure*}
\includegraphics[width=0.8\linewidth,keepaspectratio]{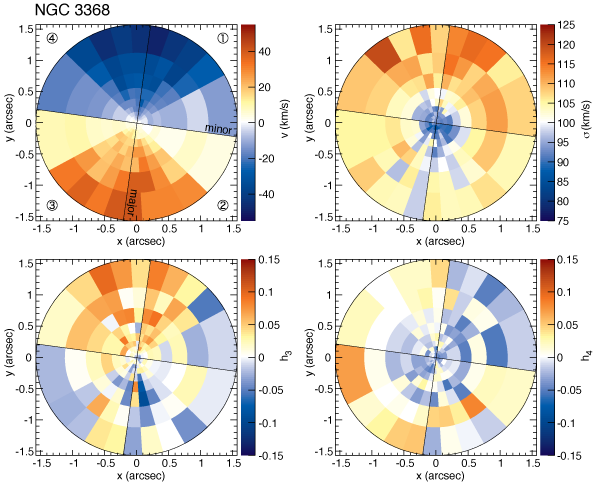}
\caption{Two-dimensional stellar kinematics ($v$, $\sigma$, $h_{3}$
and $h_{4}$) of NGC\,3368. Major axis, minor axis and the numbering of
the quadrants are indicated in the velocity map (upper
left). \label{fig:kinematics}}
\end{figure*}

For illustration purposes we fitted Gauss--Hermite polynomials to the
LOSVDs. Fig. \ref{fig:kinematics}
shows the two-dimensional fields of $v$, $\sigma$ and the higher-order
Gauss--Hermite coefficients $h_{3}$ and $h_{4}$, which quantify the
asymmetric and symmetric deviations from a Gaussian velocity profile
\citep{Gerhard-93,vanderMarel-93}. The major-axis profiles are shown
in Fig. \ref{fig:fits3368}.

The velocity field of NGC\,3368 shows a regular rotation about the
minor axis. The average, luminosity-weighted $\sigma$ within the total
SINFONI field of view is $98.5$~\kms. A central $\sigma$-drop of $7\%$
is present within the inner $\sim1$~arcsec, well inside the region of
the classical bulge component. $\sigma$-drops are not uncommon in
late-type galaxies and are usually associated with nuclear discs or
star-forming rings
(e.g. \citealt{Wozniak-03,Peletier-07,Comeron-08}). These could be
formed e.g. as a result of gas infall and subsequent star formation,
but as no change in ellipticity is found in the centre, such a disc
would have to be very close to face-on.  A $\sigma$-drop does not
imply the absence of a SMBH if the centre is dominated by the light of
a young and kinematically cold stellar population (see also the
discussion in \S 4.4). \citet{Davies-07b} observed $\sigma$-drops in a
number of strongly active galaxies. In these AGN the mass of the
central stellar component was $\sim10$ times that of the SMBH, so no
outstanding kinematic signature would be expected. Another example is
the velocity dispersion of the Milky Way, which apparently drops in
the central $100$~pc, and only rises in the inner $1-2$~pc (see Figure
9 of \citealt{Tremaine-02}).  Finally, a central $\sigma$-drop has
been found in NGC\,1399 \citep{Gebhardt-07,Lyubenova-08}, where it has
been interpreted as a signature of tangential anisotropy.

\begin{figure}
\centering
\includegraphics[width=.7\linewidth,keepaspectratio]{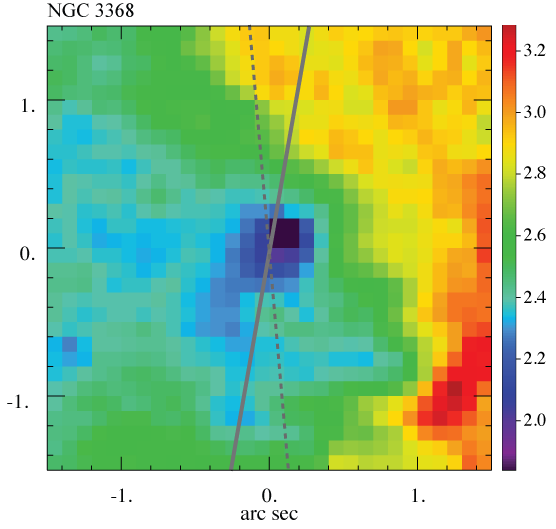}
\caption{\emph{HST} WFPC2 $B-I$ colour map of NGC\,3368. Indicated is
the major axis as a solid line (PA$=172\degr$) along with the
PA$=5\degr$ slit orientations used by \citet{Heraudeau-99} and
\citet{VegaBeltran-01}. \label{fig:extinction}}
\end{figure}

The velocity dispersion in quadrants $2$ and $3$ is smaller than in
quadrants $1$ and $4$. A possible explanation for that behaviour
could be the substantial amounts of dust in the central regions
(Fig. \ref{fig:extinction}), although the effect of the dust in the
$K$-band is relatively weak.  The \emph{HST} WFPC2 $B-I$ colour map
(Fig. \ref{fig:extinction}) shows that within the SINFONI field of
view the dust extinction is largest in quadrants $1$ and $2$. Quadrant
$4$ is moderately affected while quadrant $3$ seems to be relatively
dust-free. We will further discuss the asymmetries in \S 4.5.

In the near-IR the presence of dust should have a much smaller effect
on the kinematics than in the optical, therefore the asymmetry should
be much stronger in the kinematics measured using optical absorption
lines, if dust is responsible for the asymmetry. Longslit kinematics
($v$ and $\sigma$) at PA$=5\degr$, measured from optical spectra using
Fourier-Fitting or FCQ \citep*{Bender-94}, are available from
\citet{Heraudeau-99} and \citet{VegaBeltran-01}.  Two-dimensional
kinematics have been measured by \citet{Silchenko-03} (see also
\citealt{Moiseev-04}) with the Multi-Pupil Field Spectrograph (MPFS)
at the Russian 6~m telescope in the optical using a cross-correlation
technique. The spatial resolution of the optical data is between
$1.4$~arcsec and $3.0$~arcsec. The velocities of the
different authors are in good agreement with each other and with the
SINFONI velocities considering the different seeing values. The
optical velocity dispersions are, however, significantly larger than
those measured with SINFONI. They are on average around $130$~\kms\
for the longslit data and $\sim150$~\kms\ for the MPFS data. There are
a number of possible causes for such a discrepancy. The authors used
different correlation techniques, slightly different wavelength
regions and different templates. A difference between optical and
$K$-band $\sigma$ measurements was also found by \citet{Silge-03} for
a sample of galaxies and they suggested that this might be caused by
strong dust extinction in the optical. But weak emission lines could
also alter the absorption lines and thus the measured kinematics. As
in the SINFONI data, a velocity dispersion asymmetry is also present
in all optical datasets, as well as a velocity asymmetry. The velocity
dispersion of \citet{Moiseev-04} is enhanced in the entire region west
of the major axis, where also the majority of the dust is located
(Fig. \ref{fig:extinction}). However, when comparing the extinction
along the location of the longslits of \citet{Heraudeau-99} and
\citet{VegaBeltran-01} with the according velocity dispersion, there
seems to be no correlation. Thus it is not clear whether and in what
way dust influences the velocity dispersion in NGC\,3368.

Another explanation for the asymmetry could be lopsidedness, which is
common in late-type galaxies. Possible mechanisms which could cause
lopsidedness include minor mergers, tidal interactions and asymmetric
accretion of intergalactic gas \citep{Bournaud-05}. As the large-scale
stellar and gas velocity fields and gas distributions
\citep{Silchenko-03,Haan-08a} are rather regular, a recent merger or
collision with another galaxy seems unlikely. Accretion of gas from
the intergalactic HI cloud is a more likely scenario
\citep{Schneider-89,Silchenko-03} and could be a possible explanation
for the presence of molecular hydrogen clouds close to the centre (see
below). However, there seems to be no lopsidedness in the $K$-band
photometry, as any distortions of the isophotes can plausibly be
explained by dust.  The molecular gas distribution on the other hand
is very disordered in the central $\sim200$~pc (see below and
\citealt{Haan-08b}). Thus if the gas mass differences between
different regions of the galaxy would be large enough, they could be a
plausible explanation for the distorted stellar kinematics. However,
as shown in \S \ref{chap:discussion3368}, the molecular gas mass is
small compared to the dynamical mass and is thus unlikely to have a
significant effect on the stellar kinematics.

Central lopsidedness like an M31-like nucleus or otherwise
off-centred nuclear disc \citep{Bender-05,Jog-08} could, if the
resolution is just not high enough to resolve the disc as such, leave
certain kinematical signatures like a slightly off-centred
$\sigma$-peak or -drop. On the other hand we see velocity asymmetries
out to $r\sim20$~arcsec, which is way too large to be explained by an
M31-like nuclear disc.

In principle, the outer and inner bars could cause asymmetries in the
stellar kinematics. However, the SINFONI field of view is located well
inside the inner bar, and the only changes in velocity dispersion
associated with inner bars which have been observed are symmetric and
take place at the outer ends of inner bars \citep{deLorenzo-08}.

\subsection{Gas Kinematics in NGC\,3368}
\begin{figure*}
\includegraphics[width=0.9\linewidth,keepaspectratio]{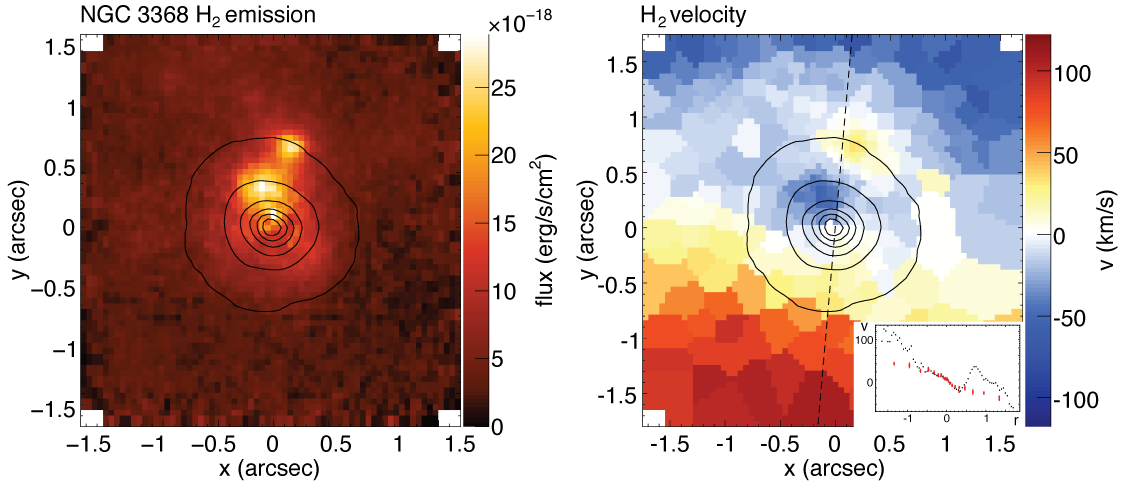}
\caption{Left panel: \hii\ (1--0S(2)) emission in the centre of
NGC\,3368. Right panel: \hii\ gas velocity. The dashed line indicates
the major axis. The small inset shows the pseudo-longslit gas velocity
profile along the major axis (black points) in comparison with the
major-axis stellar velocity (red points). The isophotes of the stellar
emission are overlaid in both panels.
 \label{fig:emission}}
\end{figure*}

In NGC\,3368 the only emission lines detected arise from molecular
hydrogen \hii. The strongest line is $1$-$0$S(1) at
$\lambda=2.1218$~$\umu$m. To determine the flux distribution and
velocity of the \hii\ gas we fitted a Gaussian convolved with a
spectrally unresolved template profile (arc line) to the
continuum-subtracted spectrum \citep{Davies-07b}. The parameters of
the Gaussian are adjusted such that they best fit the
data. Fig. \ref{fig:emission} shows the flux distribution and the
velocity field of \hii\ $1$-$0$S(1). As the S/N of the \hii\
emission is very low in some regions we binned the data using adaptive
Voronoi binning \citep{Cappellari-03} to ensure an approximately
constant S/N and thus a robust velocity measurement in each bin. This
binning scheme is different from the radial and angular binning used
to measure the stellar kinematics (see Fig. \ref{fig:kinematics}),
which is appropriate for the dynamical modelling procedure. The flux
distribution of the gas is different and more complex than the flux
distribution of the stellar light, thus it would be inappropriate to
use the same binning as for the stars for the purpose of S/N
adjustment of the gas emission. The most striking feature seen in
Fig. \ref{fig:emission} are the two clouds of \hii\ gas, located
$\sim0.36\arcsec$ and $\sim0.72\arcsec$ north of the photometric
centre. These two clouds are kinematically decoupled from the
remaining \hii\ gas distribution and seem to move in opposite
directions. Their projected sizes are approximately $25$~pc and
$20$~pc FWHM. The \hii\ distribution outside these two clouds is
relatively smooth. Its kinematic position angle, measured using the
method described in Appendix C of \citet{Krajnovic-06}, is
$\sim171\degr$ and thus agrees with the stellar kinematic position
angle. The gas velocity follows the rotation of the stars within a
radius of $\sim0.5\arcsec$; outside that radius it rotates faster,
reaching rotation velocities up to $\sim100$~\kms.

The origin of the \hii\ clouds is unclear. \hii\ emission line ratios
can help to distinguish between different excitation mechanisms like
shock heating, X-ray illumination or UV fluorescence
\citep*{Rodriguez-05}. Table \ref{tab:lineratios} shows the \hii\ line
ratios for different regions (the total SINFONI field of view and the
two \hii\ clouds). They indicate that the \hii\ gas is thermalised in
all regions of the SINFONI field of view (cf. figure 5 in
\citealt{Rodriguez-05}).

The ratio $2$-$1$S(1) 2.247~$\umu$m/$1$-$0$S(1) 2.122~$\umu$m and therefore
the vibrational excitation temperature \citep*{Reunanen-02} is smaller
in the \hii\ clouds than in the regions where \hii\ is evenly
distributed.

\begin{table*}
 \centering 
 \begin{minipage}{140mm}
  \caption{\hii\ 1--0S(1) $2.12\umu$m emission line fluxes and \hii\ line ratios.}\label{tab:lineratios}
  \begin{tabular}{ccccccccc}
  \hline
Region & 1--0S(1) ($10^{-15}$~erg~s$^{-1}$~cm$^2$) & $\frac{\mathrm{1-0S(3)}}{\mathrm{1-0S(1)}}$ & $\frac{\mathrm{1-0S(2)}}{\mathrm{1-0S(1)}}$ & $\frac{\mathrm{1-0S(0)}}{\mathrm{1-0S(1)}}$ & $\frac{\mathrm{2-1S(1)}}{\mathrm{1-0S(1)}}$ & $T_\mathrm{vib}$ (K) & $M_\mathrm{H_2}^{\mathrm{hot}}/M_\odot$ & $M_\mathrm{H_2}^{\mathrm{cold}}$ ($10^7~\mathrm{M}_\odot$)\footnote{The first value gives the cold \hii\ mass as estimated from hot \hii\ using the conversion of \citet{Mueller-06}. The second value gives the real cold gas mass after calibration against direct mass measurements from CO ($J=1-0$) at $r=1.5$~arcsec \citep{Sakamoto-99}.} \\
 \hline
cloud1 & 1.66  & 1.35 & 0.56 & 0.20  & 0.12  & 2280 &   9.13 &  2.23 | 0.21\\
cloud2 & 0.83  & 1.30 & 0.51  & 0.21  & 0.11 & 2203 &   4.57 &  1.12 | 0.10\\
total  & 21.40 & 1.69 & 0.98  & 0.26  & 0.17 & 2740 & 117.77 & 28.72 | 2.65\\
\hline
\end{tabular}
\end{minipage}
\end{table*}

\subsection{Line Strength Indices for NGC\,3368}

The stellar populations of NGC\,3368 have been analysed by
\citet{Silchenko-03} and \citet{Sarzi-05} using optical spectra. Both
found that a relatively young stellar population with a mean age of
around $3$~Gyr dominates the central region. Towards larger radii
\citet{Silchenko-03} found a strong increase in age.

\begin{figure*}
\includegraphics[width=\linewidth,keepaspectratio]{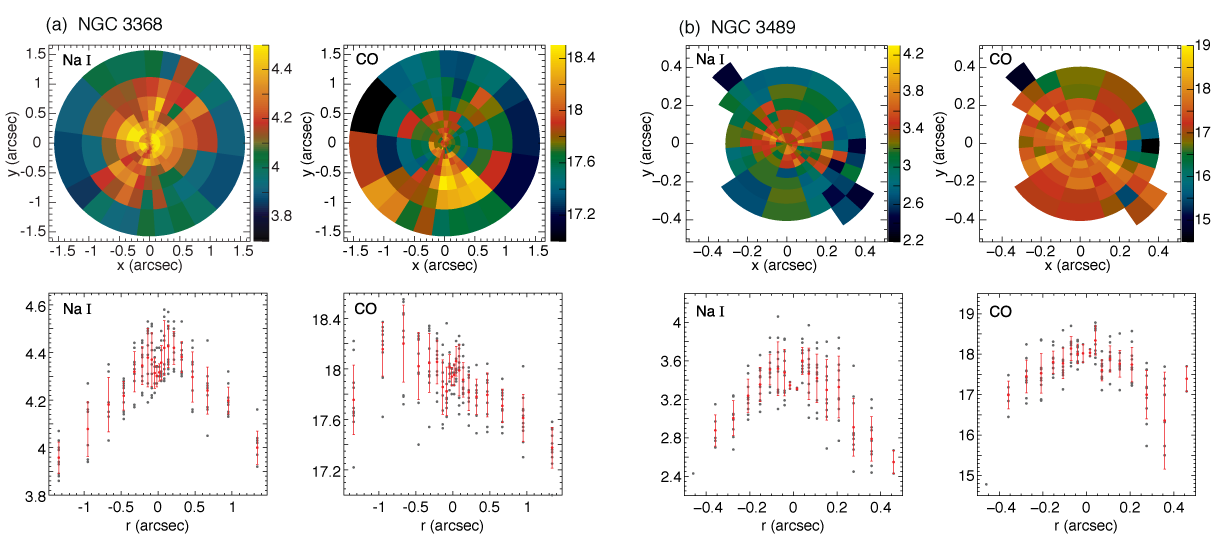}
\caption{Near-IR line indices \nai\ and CO in {\AA} for (a)
NGC\,3368 and (b) NGC\,3489. The upper row shows the two-dimensional
index fields, with the position-index diagrams in the bottom
row. Individual values are plotted in grey, while overplotted in red
are the mean indices and their rms for bins belonging to
semicircles in the receding half of the galaxy (negative radius $r$) or
the approaching half (positive radius $r$).\label{fig:indices}}
\end{figure*}

Fig. \ref{fig:indices} shows the near-IR line indices \nai\ and CO
measured in the same way as in \citet{Nowak-08}, using the definitions
of \citet*{Silva-08}. The average values inside the SINFONI field of
view are listed in Table \ref{tab:index}. They differ significantly
from the relations between \nai\ or CO and $\sigma$ found by
\citet{Silva-08} for early-type galaxies in the Fornax cluster which
may be, as in the case of Fornax\,A, probably due to the relatively
young age of the stellar population. Younger populations seem to have
larger \nai\ at equal $\sigma$ than old stellar populations in the
galaxy samples of \citet{Silva-08} and \citet{Cesetti-08}. However, no
such trend is obvious for the CO index, so the difference seen here
could be due to other aspects like metallicity or galaxy formation
history. The measured average indices have values which are quite
similar to those found in the centre of Fornax\,A \citep{Nowak-08},
which could indicate that the stellar populations are quite similar in
terms of age and metallicity. However, the interpretation of \nai,
\cai\ and \fei\ must be done bearing in mind that these
features always include significant contributions from other elements
\citep{Silva-08}.

\begin{table}
 \centering
  \caption{Mean near-IR line strength indices in {\AA} of NGC\,3368 ($3\times3$~arcsec aperture) and NGC\,3489 ($0.8\times0.8$~arcsec aperture). The corresponding rms is given in brackets.}\label{tab:index}
  \begin{tabular}{ccc}\\
  \hline
   &  NGC\,3368 & NGC\,3489\\
 \hline
\nai & 4.30 (0.16) & 3.30 (0.36)\\
\cai & 2.40 (0.29) & 1.53 (0.50)\\
\feia & 1.55 (0.07) & 1.25 (0.23)\\
\feib & 0.91 (0.10) & 0.76 (0.21)\\
CO   & 17.90 (0.27) & 17.70 (0.72)\\
\hline
\end{tabular}

\end{table}

The radial distribution of the line indices (Fig. \ref{fig:indices})
shows a slight asymmetry, similar to the kinematics. The two quadrants
with the smaller $\sigma$ have larger CO EWs and smaller \nai\ EWs
than the other two quadrants. In addition there seems to be a strong
negative gradient in \nai\ and a moderately strong negative gradient
in CO. \cai\ and \fei\ are approximately constant with radius. A small
central drop is present in most indices, which could indicate the
presence of weak nuclear activity (see \citealt{Davies-07b}).

\subsection{Stellar Kinematics in NGC\,3489}

The NGC\,3489 data were binned in the same way as the NGC\,3368 data,
with identical angular bins and nine somewhat smaller radial bins due to
the higher spatial resolution. A position angle of $71\degr$ was
used. The mean S/N is $70$, and a smoothing parameter of
$\alpha=8$ was used. As kinematic template stars we chose
four K and M giants with an intrinsic CO equivalent
width in the range 13--15~{\AA}.

\begin{figure*}
\includegraphics[width=0.8\linewidth,keepaspectratio]{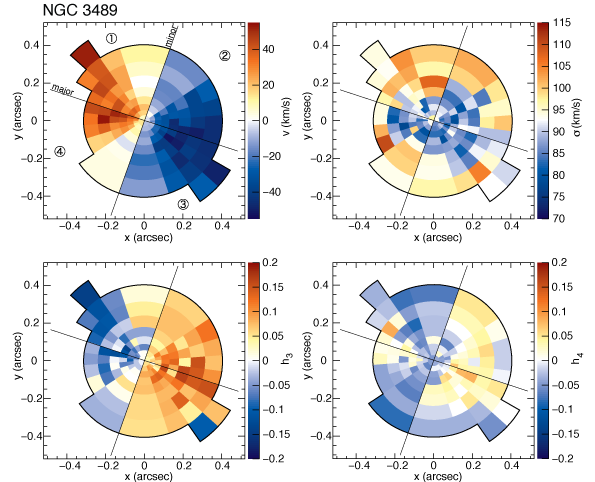}
\caption{Two-dimensional stellar kinematics ($v$, $\sigma$, $h_{3}$
and $h_{4}$) of NGC\,3489. Major axis, minor axis and the numbering of
the quadrants are indicated in the velocity map (upper
left).\label{fig:kinematics3489}}
\end{figure*}

Fig. \ref{fig:kinematics3489} shows the $v$, $\sigma$, $h_{3}$ and
$h_{4}$ maps of NGC\,3489. The kinematics is similar to that of
NGC\,3368 in some aspects. It is clearly rotating about the minor
axis, though stronger than NGC\,3368. The velocity dispersion also
drops towards the centre by around $4\%$, but then has a tiny peak in
the central bins. The average $\sigma$ in the total
$0.8\times0.8$~arcsec$^2$ field of view is $91$~\kms. The $h_{3}$
values clearly anticorrelate with $v$. $h_{4}$ is on average small and
negative. Along the major axis it is positive in the outer bins and
negative in the inner bins. No asymmetry is present in $\sigma$, but
the velocity is slightly asymmetric. It increases strongly on the
receding side and then remains approximately constant at
$r>0.05$~arcsec, whereas on the approaching side the slope is less
steep and an approximately constant velocity is reached much further
out at $r>0.2$~arcsec. The major-axis profiles are shown in
Fig. \ref{fig:fits3489}. Note that the central velocity bin is omitted
in that plot. Despite the presence of strong dust features in optical
images, the kinematics is in comparatively good agreement with the 2D
SAURON \citep{Emsellem-04} and OASIS \citep{McDermid-06} kinematics,
though due to the high spatial resolution and the very small field of
view of our data a direct comparison with seeing-limited data is not
easy. The SINFONI velocity field seems to be fully consistent with the
optical velocity. The average SINFONI velocity dispersion is smaller
than the central SAURON and OASIS $\sigma$. The central SAURON and
OASIS $h_{4}$ is significantly larger than the SINFONI values, and the
anticorrelation of $h_{3}$ and $v$ seems to be less strong in general
and essentially non-existent in the central arcsecond of the OASIS
data.

Longslit kinematics is available from \citet*{Caon-00}, who
used a cross-correlation technique to determine $v$ and
$\sigma$. Their $\sigma$ is much larger ($117$~\kms\ in the central
pixels of the major-axis longslit), which could be due to their
cross-correlation technique or template mismatch. In addition the
$\sigma$-profile seems to be slightly asymmetric, but only at large
radii, where the errors are large.

\subsection{Line Strength Indices for NGC\,3489}
The stellar populations of NGC\,3489 in the centre have been analysed
by \citet{Sarzi-05} using optical \emph{HST} STIS longslit spectra and
by \citet{McDermid-06} using OASIS integral-field data.
\citet{Sarzi-05} obtained a mean age of about $3.1$~Gyr in the central
$0.2\times0.25$~arcsec by fitting stellar population synthesis models
to the spectra, assuming solar metallicity. \citet{McDermid-06}
obtained a mean age of $1.7$~Gyr in the central $8\times10$~arcsec$^2$
with an age gradient down to $\sim1$~Gyr towards the centre from the
analysis of Lick indices. These two values are more or less in
agreement when taking into account the measurement errors and that the
\citet{Sarzi-05} value would decrease when considering the metallicity
increase to supersolar values in the centre measured by
\citet{McDermid-06}. Another possibility is that the central
$\sim0.2$~arcsec, which are unresolved by \citet{McDermid-06}, contain
an older stellar population.

The near-IR absorption-line indices \nai\ and CO are shown in
Fig. \ref{fig:indices}b. They seem to be, like the stellar kinematics,
axisymmetric. As in the case of NGC\,3368 there is a clear negative
gradient in both indices, which could mean an age or a metallicity
gradient, or a combination of both. The other indices, \cai\ and \fei,
are largely constant. The average line strength indices within the
$0.8\times0.8$~arcsec$^2$ field of view are given in Table
\ref{tab:index}. The average CO line strength is very similar to the
value found in NGC\,3368.  All other measured indices are slightly
smaller than in NGC\,3368. This seems to be generally in agreement
with the results of \citet{Sarzi-05}, who found similar mean ages and
populations in both galaxies. A small central drop is present only in
\nai, implying that nuclear activity must be extremely weak or
absent.

\section{Dynamical Modelling of NGC\,3368}

For the dynamical modelling we make use of the \citet{Schwarzschild-79}
orbit superposition technique: first the gravitational potential of
the galaxy is calculated from the stellar luminosity density $\nu$ and
trial values for the black-hole mass $M_{\bullet}$ and the
mass-to-light ratio $\Upsilon$. Then an orbit library is generated for
this potential and a weighted orbit superposition is constructed such
that it matches the observational constraints.  Finally everything is
repeated for other potentials until the appropriate parameter space in
$M_{\bullet}$ and $\Upsilon$ is systematically sampled. The
best-fitting parameters then follow from a $\chi^2$-analysis. The
deprojected luminosity density is a boundary condition and thus is
exactly reproduced, while the LOSVDs are fitted. 

\subsection{Construction of the Stellar Luminosity Profile} 

For dynamical modelling purposes, we need an appropriate
surface-brightness profile and an appropriate ellipticity profile,
along with an assumption of axisymmetry.  While simply using the
results of ellipse-fitting may be valid for an elliptical galaxy,
where the approximation that the galaxy is a set of nested,
axisymmetric ellipsoids with variable axis ratio but the same position
angle is often valid, a system like NGC\,3368, with two bars, dust
lanes, and spiral arms, is clearly more complicated. Such a complex
structure also makes it important to allow a $\Upsilon$ gradient in
order to account for stellar population changes. This can be
conveniently approximated by using more than one component, where
each component has its own $\Upsilon$.

We model the luminosity distribution of NGC\,3368 as the combination
of \textit{two} axisymmetric components: a disc with fixed (observed)
ellipticity = 0.37, which by design includes both inner and outer bars
\textit{and} the discy pseudobulge; and a central ``classical'' bulge
of variable (but low) ellipticity.  Thus, we assume that the bars can,
to first order, be azimuthally ``averaged away.''

The surface brightness profile of the disc component is \textit{not}
assumed to be a simple exponential.  Instead, it is the
\textit{observed} surface brightness profile of the entire galaxy
outside the classical bulge, out to $r=130$~arcsec, along with an inward
extrapolation to $r=0$. We base this profile on ellipse fits with
fixed ellipticity and position angle ($\epsilon=0.37$, PA $=
172\degr$) to the $K$-band image of \citet{Knapen-03}, with the inner
$r<3.7$~arcsec based on the exponential component of our inner
bulge-disc decomposition (Fig.~\ref{fig:inner-decomp-n3368}).
(Comparison of profiles from the dust-corrected NICMOS2 image and the
Knapen et al.\ image shows that seeing affects the latter only for $r
< 2$~arcsec, which is already within the region where the classical
bulge affects the profile.)  Inspection of
both this profile and a similar fixed-ellipse profile from the NICMOS2
image shows that the classical bulge begins to affect the profile only
for $r < 3.7$~arcsec. Consequently, the disc profile for $r <
3.7$~arcsec is the inward extrapolation of the exponential component
from our inner S\'ersic+exponential decomposition ($\mu_0 = 12.75$, $h =
5.28$~arcsec; Fig.~\ref{fig:inner-decomp-n3368}).

To generate the profile of the classical bulge, we assume, following the
inner decomposition discussed above, that the light in the inner
$r<8$~arcsec is the combination of an inner exponential and the
classical bulge (Fig. \ref{fig:inner-decomp-n3368}).  We generated a
model image with the same size as the NICMOS2, containing a 2D
exponential model for the inner disc, which we subtracted from the
NICMOS2 image. The residual image is assumed to contain light from the
classical bulge only; we then fit ellipses to this image. This allows
for possible variations in the classical bulge's ellipticity and,
perhaps more importantly, uses the observed surface brightness profile
at the smallest radii, rather than an analytic fit. Finally, we generate
an extension of this bulge profile out to the same outermost radius as
the disc profile (i.e., well outside the NICMOS2 image) by fitting a
S\'ersic function to the classical-bulge profile, and assuming a
constant ellipticity of 0 and the same PA as for the outer disc at large
radii.

The program of \citet{Magorrian-99} was used for the deprojection
assuming that all components are axisymmetric. Both components, the
disc and the classical bulge, were deprojected for an inclination
$i=53\degr$ as obtained from the photometry (see \S 2.3.4), and for a
few nearby values between $52\degr$ and $55\degr$. No shape penalty
was applied. The simplest assumption for the form of the stellar mass
density $\rho_*$ is then
$\rho_{*}=\Upsilon_{\mathrm{bulge}}\cdot\nu_{\mathrm{bulge}}+\Upsilon_{\mathrm{disc}}\cdot\nu_{\mathrm{disc}}$,
where $\nu$ is the deprojected luminosity density and
$\Upsilon_\mathrm{bulge}$ and $\Upsilon_\mathrm{disc}$ two constants
to be determined \citep{Davies-06}. The assumption of a constant
$\Upsilon$ is approximately true for the central part of the galaxy
where we have kinematic data (dark matter does not play a significant
role).

As a further test, we deprojected the two components from the
global bulge-disc decomposition (i.e. the outer exponential disc and a
S\'{e}rsic fit to the photometric bulge region, as shown in
Fig. \ref{fig:global-decomp-n3368}). The resulting shape of the
luminosity profile is very similar to the profile obtained from the
inner bulge-disc decomposition of the photometric bulge. The global
profile is offset to smaller luminosities, as the global decomposition
does not fully account for the light in the classical bulge
component. This means that the SMBH mass estimates
would be larger for models based on the global bulge-disc
decomposition (for constant $\Upsilon$) compared to the mass estimate derived
from models using the inner bulge-disc decomposition.

\subsection{Dynamical Models}

As in \citet{Nowak-07,Nowak-08} we use an axisymmetric code
\citep{Richstone-88,Gebhardt-00c,Gebhardt-03,Thomas-04} to determine
the mass of the SMBH in NGC\,3368. This method has been successfully
tested on the maser galaxy NGC\,4258 in \citet{Siopis-09}, who
obtained the same mass for the black hole as determined from maser
emission.  

We allow for different mass-to-light ratios in the classical
bulge region and the region further out, respectively. Radial changes in
the mass-to-light ratio can bias the derived BH mass if not taken into
account properly \citep[e.g.,][]{Gebhardt-09}.

Using an axisymmetric code for a barred and therefore
obviously non-axisymmetric galaxy might be
debatable. \citet{Thomas-07} have found that axisymmetric dynamical
models of extremely triaxial/prolate systems are in danger of
underestimating the luminous mass in the centre. Since BH mass and
central luminous mass are partly degenerate, this could result in an
overestimate of the BH mass. For two-component models the situation is
even more complex. If the triaxiality only affects the outer
$\Upsilon_\mathrm{disc}$ (as in the case of NGC 3368), then a
corresponding underestimation of the outer $\Upsilon_\mathrm{disc}$
could translate into an overestimation of the inner
$\Upsilon_\mathrm{bulge}$, which in turn would imply a bias towards
low BH masses. Detailed numerical simulations of barred galaxies are
required to investigate such possible biases. 

However, in this case the axisymmetric models can be justified as we
only model the central part of the galaxy. Near the SMBH the potential
is intrinsically spherical and strong non-axisymmetries are
unlikely. Also, there is little evidence that non-axisymmetric bar
orbits dominate the observed region in projection (see \S 4.5).

We use only the SINFONI data for the modelling. The four quadrants are
modelled separately in order to assess the influence of deviations
from axisymmetry. These four independent measurements of the SMBH mass
should agree within the observational errors, if the galaxy is
axisymmetric. If not, then the systematic differences
from quadrant to quadrant provide an estimate for the systematic
errors introduced by assuming axial symmetry.

Since the observed ellipticity of the classical bulge is affected by
strong dust lanes, the ellipticity is slightly uncertain (see \S
2.3.1). We ran dynamical models for two different deprojections,
one obtained for a bulge ellipticity $\epsilon=0.0$ and the other for
$\epsilon=0.1$. The models yield the same mass-to-light ratios and
black hole masses and in the following we only discuss the case
$\epsilon=0.0$.

In order to find out whether the results depend on the assumed
inclination of the galaxy, we run models for four different
inclinations around the most likely value of $53\degr$.

We do not apply regularisation to our models, because
the exact amount of regularisation is difficult to determine due to
the lack of realistic, analytical models of disc galaxies with black
holes.

The SINFONI observations mainly cover the classical bulge region, so
the disc $\Upsilon$ can only be weakly constrained. It could be better
constrained if we included other kinematic data extending further out,
but this has several disadvantages. The inconsistencies between the
SINFONI and the optical measurements from the literature mean that the
models could have difficulties fitting the different kinematic datasets
reasonably well at the same time.  In addition, the non-axisymmetries
due to the two bars would be more noticeable at large radii, and a
dark halo would become important.

\subsection{Results}
\begin{figure*}
\includegraphics[width=\linewidth,keepaspectratio]{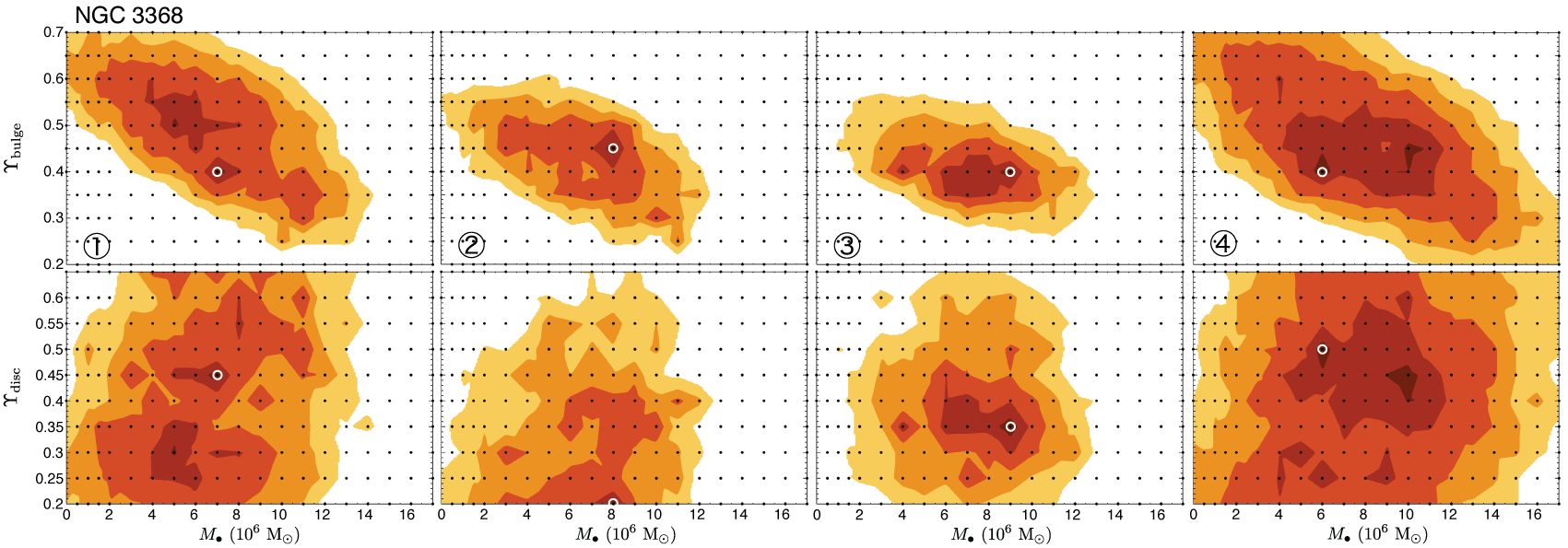}
\caption{Models for NGC\,3368 calculated for the quadrants $1$-$4$, an
inclination $i=53\degr$ and a classical bulge ellipticity
$\epsilon=0.0$. For each quadrant
$\Delta\chi^{2}_{0}=\chi^{2}-\chi^{2}_{\mathrm{min}}$ is plotted as a
function of the black hole mass $M_{\bullet}$ and the $K$-band
mass-to-light ratios $\Upsilon_{\mathrm{bulge}}$ (top row) and
$\Upsilon_{\mathrm{disc}}$ (bottom row). The coloured regions are the
$1$-$5\sigma$ confidence intervals for two degrees of freedom. Each
model that was calculated is marked as a black dot; the best-fitting
model is encircled by a white ring.\label{fig:models}}
\end{figure*}

\begin{figure*}
\includegraphics[width=0.8\linewidth,keepaspectratio]{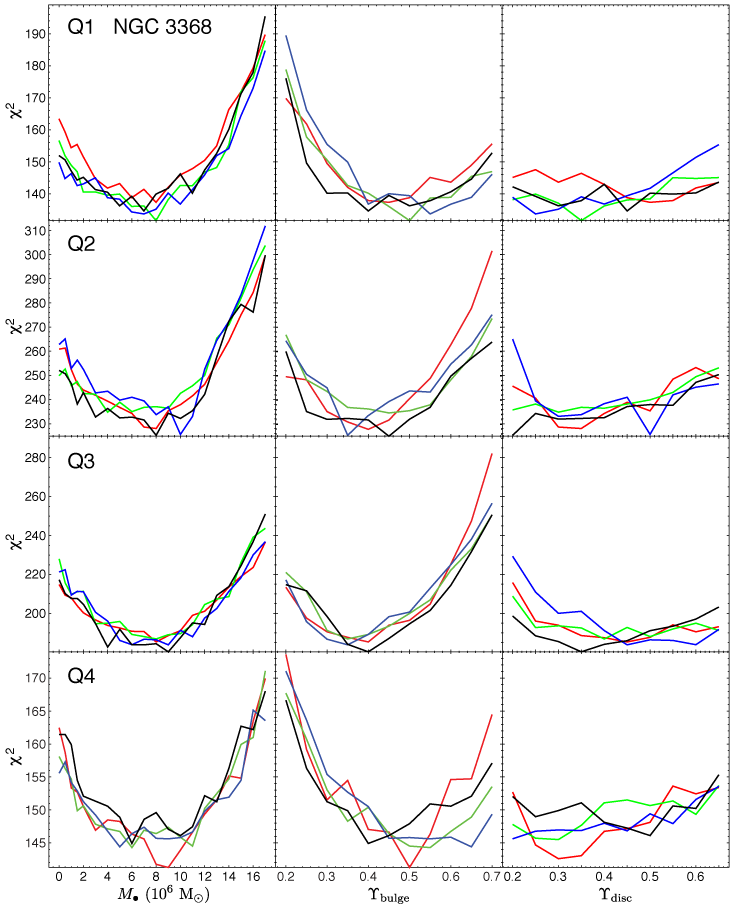}
\caption{$\chi^2$ as a function of $M_\bullet$ (left column,
marginalised over $\Upsilon_{\mathrm{bulge}}$ and
$\Upsilon_{\mathrm{disc}}$), $\Upsilon_{\mathrm{bulge}}$ (middle
column, marginalised over $M_\bullet$ and $\Upsilon_{\mathrm{disc}}$)
and $\Upsilon_{\mathrm{disc}}$ (right column, marginalised over
$M_\bullet$ and $\Upsilon_{\mathrm{bulge}}$) for NGC\,3368. Different
inclinations are indicated by different colours (red: $i=52\degr$,
black: $i=53\degr$, green: $i=54\degr$, blue: $i=55\degr$). \label{fig:chi2_3368}}
\end{figure*}

The results for $i=53\degr$ and $\epsilon=0.0$ are shown in
Fig. \ref{fig:models} ($\Delta\chi^2$ as a function of $M_\bullet$ and
the $\Upsilon$ of one component, marginalised over the other
component's $\Upsilon$), and for all inclinations in
Fig. \ref{fig:chi2_3368} (total $\chi^2$ as a function of one of the
three parameters $M_\bullet$, $\Upsilon_{\mathrm{bulge}}$,
$\Upsilon_{\mathrm{disc}}$, marginalised over the other two
parameters). The best-fitting values with $3\sigma$ errors are listed
in Table \ref{tab:results} for all four inclinations. Each
quadrant provides an independent measurement of $M_\bullet$ if
deviations from axisymmetry do not play a big role. This seems to be
the case, as the resulting best-fitting values for $M_{\bullet}$,
$\Upsilon_{\mathrm{bulge}}$ and $\Upsilon_{\mathrm{disc}}$ agree very
well within $\la2\sigma$ between the four quadrants.  The mean black
hole mass for the four quadrants ($i=52-55^\circ$) is $\langle
M_{\bullet}\rangle=7.5\times10^{6}$~M$_{\odot}$
($\mathrm{rms}(M_{\bullet})=1.5\times10^{6}$~M$_{\odot}$). Note
that the average of all $3$~$\sigma$ errors given in Table
\ref{tab:results} derived from the $\chi^2$ analysis divided by three
is $1.6\times10^6$~M$_\odot$ and thus approximately equal to the rms
error.

The resulting black hole mass does not depend much on the particular
choice of the mass-to-light ratio of the disc
$\Upsilon_{\mathrm{disc}}$, but decreases for increasing
$\Upsilon_{\mathrm{bulge}}$. As shown in Fig. \ref{fig:chi2_3368}, the
results also do not change systematically with the inclination. This
shows that the inclination cannot be constrained better by dynamical
modelling than by a thorough analysis of photometric data, and that a
very precise knowledge of the inclination is not necessary for
dynamical modelling purposes. 

\subsection{Evidence for a black hole in NGC 3368}
\begin{figure*}
\includegraphics[width=0.8\linewidth,keepaspectratio]{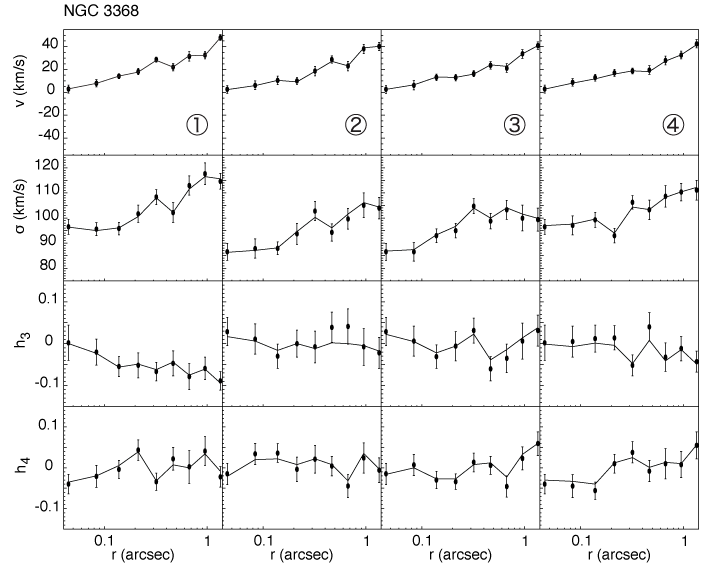}
\caption{The major axis kinematics of NGC\,3368 is shown as black
points for quadrants 1 to 4. Overplotted is the fit of the best
model with black hole.
\label{fig:fits3368}}
\end{figure*}

\begin{figure*}
\includegraphics[width=0.7\linewidth,keepaspectratio]{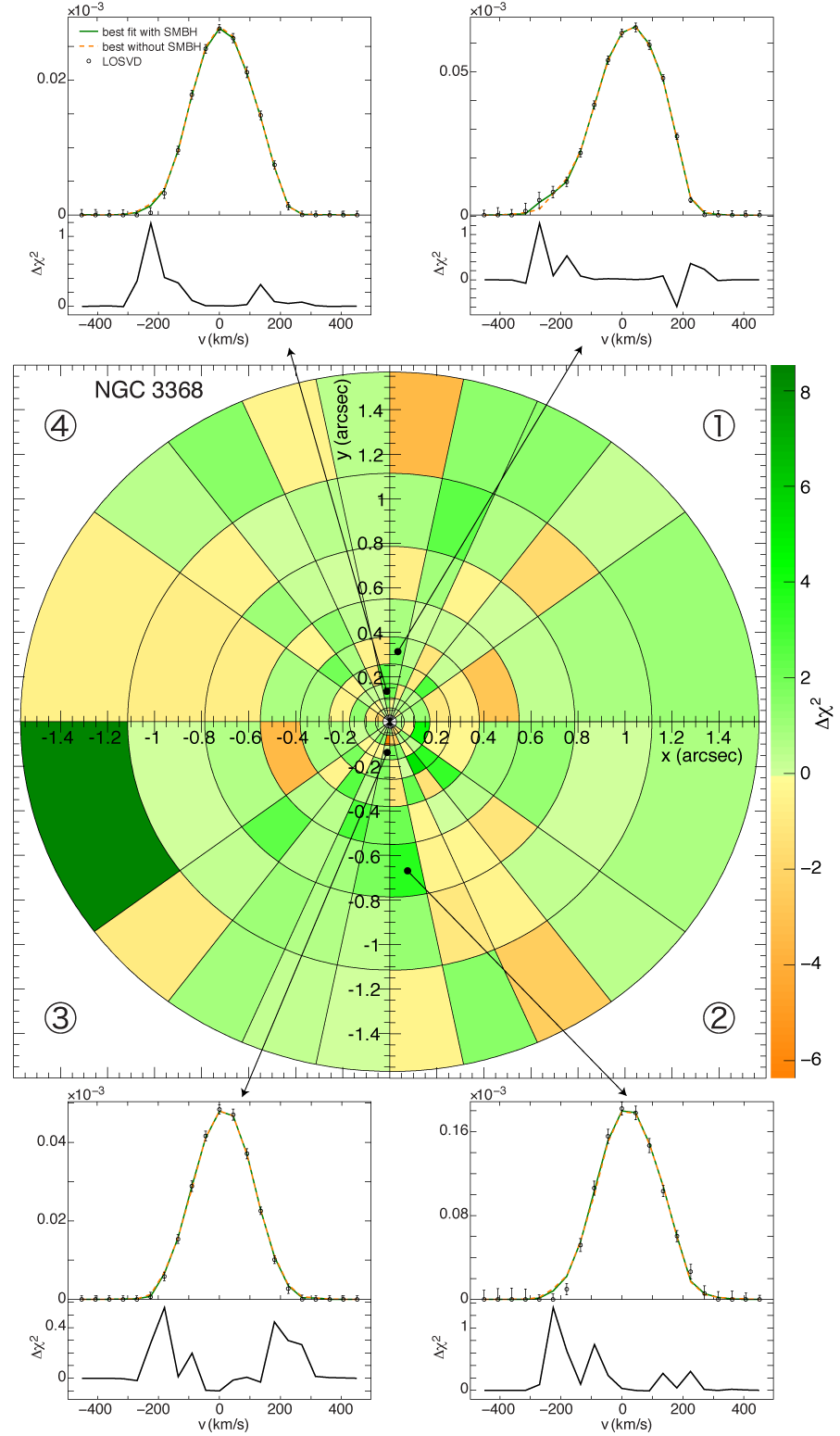}
\caption{$\chi^{2}$ difference between the best-fitting model without
black hole and the best-fitting model with black hole
($\Delta\chi^{2}=\sum_{i}\Delta\chi_{i}^{2}=\sum_{i=1}^{21}(\chi^{2}_{i,\mathrm{noBH}}-\chi^{2}_{i,\mathrm{BH}})$
over all 21 velocity bins) for all LOSVDs of the four quadrants of
NGC\,3368. Bins where the model with black hole fits the LOSVD better
are plotted in green, the others in orange. For four bins along the
major axis the LOSVDs (open circles with error bars, normalized as in
\citealt{Gebhardt-00c}) and both fits (with black hole, full green
line and without black hole, dashed orange line) are shown with the
corresponding $\Delta\chi_i^2$ plotted below the
LOSVDs. \label{fig:green3368}}
\end{figure*}

$v$, $\sigma$, $h_3$ and $h_4$ of the best model at $i=53\degr$ and
$\epsilon=0.0$ (major axis) are shown in Fig. \ref{fig:fits3368} for
all quadrants. The corresponding best fit without black hole would be
very similar, which is why we choose to plot the differences in
$\chi^2$ instead (see below and Fig. \ref{fig:green3368}). The
similarity between the black-hole model and the one without black hole
in the lower order velocity moments, as well as the $\sigma$-drop
towards the centre raise the question where the dynamical evidence for
the black hole comes from. 

\begin{figure}
\includegraphics[width=0.95\linewidth,keepaspectratio]{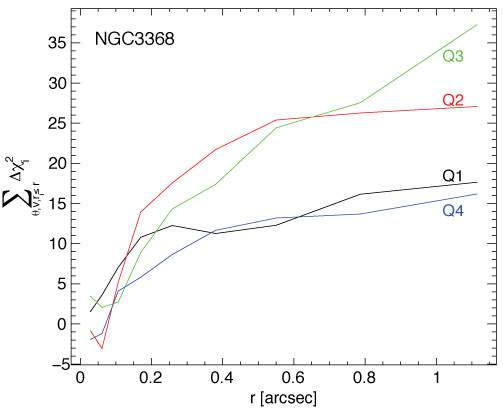}
\caption{$\chi^{2}$ difference between the best-fitting model without
black hole and the best-fitting model with black hole summed over all
angular and velocity bins as a function of distance from the black
hole for all four quadrants of NGC\,3368. \label{fig:radchi2_3368}}
\end{figure}

Fig. \ref{fig:green3368} shows the $\chi^2$ difference between the
best-fitting model without a black hole and the best-fitting model
with a black hole for all LOSVDs in all four quadrants. The fit with
black hole is generally better in $132$ of the $180$ bins. The largest
$\chi^2$ differences appear in the LOSVD wings, where the model with
black hole gives less residuals. As a different illustration of
what is shown in Fig. \ref{fig:green3368}, Fig. \ref{fig:radchi2_3368}
shows the increase in $\Delta\chi^2$ summed over all angular and
velocity bins as a function of radius. Thus for each quadrant
five (angles) times $21$ (velocities) $\Delta\chi^2$ values
were added at each radius. The largest $\Delta\chi^2$ increase
occurs in the central $\sim0.2-0.3$~arcsec
(i.e. $\sim4r_\mathrm{SoI}$). At larger radii the $\Delta\chi^2$
increase is less strong. $\Upsilon_{\mathrm{bulge}}$ is large for the
best models without black hole, which can worsen the fit in the outer
data regions. Therefore it is not surprising that improvements of the
fit appear at all radii. The total $\Delta\chi^2$, summed over all
LOSVDs, between the best-fitting model without a black hole and the
best-fitting model with a black hole is given in the last column of
Table \ref{tab:results}.

The asymmetry of the data is reflected in the error bars. For
quadrants $1$ and $4$, which are the quadrants with the higher
velocity dispersion, the error bars are much larger than for quadrants
$2$ and $3$. In $2$ and $3$ the no-black hole-solution is excluded by
$\ga5\sigma$, whereas for $1$ and $4$ it is only excluded by
$\sim3.6\sigma$.

\begin{table*}
 \centering
 \begin{minipage}{140mm}
  \caption{Resulting black hole masses and $K$-band mass-to-light ratios $\Upsilon_{\mathrm{bulge}}$ and $\Upsilon_{\mathrm{disc}}$ of NGC\,3368. The lower and upper $3\sigma$ limits are given in brackets. The total $\chi^2$ of the best model with black hole and the $\chi^2$ difference between the best model without black hole and the best model with black hole are given in the last two columns.}\label{tab:results}
  \begin{tabular}{lllllll}
  \hline
Inclination & Quadrant & $M_\bullet$ [$10^6$~M$_\odot$] & $\Upsilon_\mathrm{bulge}$ & $\Upsilon_\mathrm{disc}$ & $\chi_{\mathrm{min}}^{2}$ & $\Delta\chi^2_\mathrm{noBH-BH}$\\
 \hline
$52\degr$ & 1 & 8.0 (3.0, 11.0) & 0.45 (0.30, 0.60) & 0.50 (0.20, 0.65) & 137.353 & 26.093 \\
          & 2 & 8.0 (4.0, 10.0) & 0.40 (0.30, 0.45) & 0.35 (0.30, 0.50) & 228.051 & 32.828 \\
          & 3 & 8.0 (3.0, 10.0) & 0.40 (0.25, 0.50) & 0.45 (0.25, 0.65) & 185.363 & 29.471 \\
          & 4 & 9.0 (1.0, 13.0) & 0.50 (0.30, 0.55) & 0.30 (0.25, 0.65) & 141.261 & 21.207 \\ 
$53\degr$ & 1 & 7.0 (1.5, 11.0) & 0.40 (0.30, 0.65) & 0.45 (0.20, 0.65) & 134.681 & 17.285\\
          & 2 & 8.0 (3.0, 10.0) & 0.45 (0.25, 0.50) & 0.20 (0.20, 0.40) & 225.221 & 26.828\\
          & 3 & 9.0 (4.0, 10.0) & 0.40 (0.35, 0.45) & 0.35 (0.25, 0.50) & 180.291 & 36.955\\
          & 4 & 6.0 (1.5, 14.0) & 0.40 (0.25, 0.65) & 0.50 (0.20, 0.65) & 144.920 & 16.538\\ 
$54\degr$ & 1 & 8.0 (2.0, 11.0) & 0.50 (0.35, 0.60) & 0.35 (0.20, 0.50) & 131.651 & 25.002\\
          & 2 & 4.0 (1.0, 11.0) & 0.45 (0.30, 0.55) & 0.30 (0.20, 0.55) & 234.792 & 15.106\\
          & 3 & 8.0 (3.0, 11.0) & 0.35 (0.30, 0.45) & 0.40 (0.25, 0.65) & 186.925 & 41.063\\
          & 4 & 6.0 (1.0, 14.0) & 0.55 (0.30, 0.70) & 0.30 (0.20, 0.65) & 144.291 & 13.807\\ 
$55\degr$ & 1 & 7.0 (0.5, 12.0)  & 0.55 (0.40, 0.65) & 0.25 (0.20, 0.50) & 133.730 & 16.099 \\
          & 2 & 10.0 (8.0, 11.0) & 0.35 (0.35, 0.40) & 0.50 (0.30, 0.50) & 225.629 & 37.159 \\
          & 3 & 9.0 (4.0, 11.0)  & 0.35 (0.25, 0.40) & 0.45 (0.40, 0.65) & 183.933 & 37.418 \\
          & 4 & 5.0 (0.0, 15.0)  & 0.65 (0.30, 0.70) & 0.20 (0.20, 0.65) & 144.406 & 11.170 \\
\hline
\end{tabular}
\end{minipage}
\end{table*}

\subsection{Discussion}\label{chap:discussion3368}
Upper limits for $M_\bullet$ in NGC\,3368 have been measured from central
emission-line widths by \citet{Sarzi-02} and \citet{Beifiori-08}, who
obtain $2.7\times10^7$~M$_\odot$ (stellar potential included) and
$4.8\times10^7$~M$_\odot$ (stellar potential not included),
respectively. Based on the $M_\bullet$-$\sigma$ relation of
\citet{Tremaine-02}, a black hole with a mass between $8\times10^6$ and
$2.5\times10^7$~M$_\odot$ would have been expected, depending on which
$\sigma$ measurement is used. From the relation between $M_\bullet$
and $K$-band luminosity \citep{MarconiHunt-03} we would have expected a very
high black-hole mass of $9.2\times10^{7}$~M$_\odot$ if it correlates
with the total (photometric) bulge luminosity ($M_K=-23.42$), or a very
small mass of only $1.5\times10^{6}$~M$_\odot$ if it correlates with
the classical bulge luminosity ($M_K = -19.48$).


The stellar mass within the sphere of influence
$r_\mathrm{SoI}=GM_\bullet/\sigma^2\approx0.07$~arcsec is
$\approx5.7\times10^6$~M$_\odot$. If the best-fitting mass for the
black hole of $M_{\bullet}=7.5\times10^6$~M$_{\odot}$ were entirely
composed of stars, the mass-to-light ratio
$\langle\Upsilon_{\mathrm{bulge}}\rangle\approx0.41$ within
$r_\mathrm{SoI}$ would increase to $0.95$.  This would be typical for
an older stellar population ($\sim4$--$7$~Gyr for a Salpeter IMF and
$\sim10$--$11$~Gyr for a Kroupa IMF at solar metallicity, using the
models of \citealt{Maraston-98,Maraston-05}). However, this would
strongly conflict with \citet{Sarzi-05}, who find that a $1$~Gyr old
population dominates, with some contributions from older and younger
populations, resulting in a mean age of 3~Gyr.

As mentioned in \S 4.2, bar orbits crossing the centre could in
principle produce deviations from axisymmetry. Non-axisymmetric
structures such as a prolate central structure can be recognised in
the kinematics as a low-$\sigma$, high-$h_4$ region if seen edge-on,
or as a high-$\sigma$, low-$h_4$ region if seen face-on
\citep{Thomas-07}. This could also bias the reconstructed
masses. Strictly speaking, this is only valid for N-body ellipticals with a
central prolate structure. Simulations of bars by \citet{Bureau-05}
resulted in variable $\sigma$ and $h_4$, depending on the projection
of the bar. However, these variations were always symmetric with
respect to the bar. Thus we would expect symmetry between quadrants
where the bar appears. This is not the case, therefore it is unlikely
that bar orbits in projection disturb the central kinematics
significantly.

Dust could in principle influence the kinematics and produce
distortions or asymmetries, though no clear correlation with the dust
distribution could be found in \S 3.2. According to
\citet{Baes-03b}, however, dust attenuation should not affect moderately
inclined galaxies significantly.

\begin{figure}
\includegraphics[width=\linewidth,keepaspectratio]{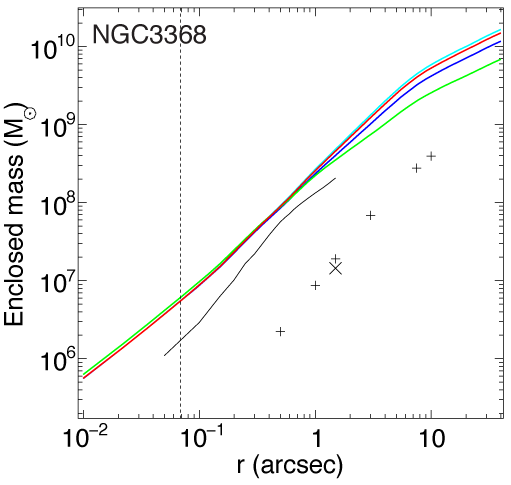}
\caption{Enclosed mass as a function of radius in NGC\,3368 for
molecular mass (black line) and stellar dynamical mass. For the
latter, we plot values estimated from our modelling of the individual
quadrants (Q1: red, Q2: green, Q3: blue, Q4: cyan). The molecular gas
mass is estimated from the hot \hii\ emission in the SINFONI data,
converted into a cold gas mass using \citet{Mueller-06}. The cold gas
mass distribution directly measured from CO emission by
\citet{Sakamoto-99} is marked by crosses and the cold gas mass from
BIMA SONG \citep{Helfer-03} is marked as an `$\times$'. The cold gas
mass derived using the hot \hii\ conversion is clearly overestimated
compared to the mass derived from CO. The vertical dashed line marks
approximately the radius of the sphere of influence, defined as $r_\mathrm{SoI}=GM_\bullet/\sigma^2$.
\label{fig:gasmass}}
\end{figure}

The mass of the hot \hii\ in the clouds and in the total field of view
was estimated via
$M_{\mathrm{H}_{2}}=5.0875\times10^{13}D^{2}I_{1-0\mathrm{S}(1)}$
\citep{Rodriguez-05} and the total cold gas mass via
$\log\left(L_{1-0\mathrm{S}(1)}
M_\mathrm{H_2}^{\mathrm{cold}}\right)=-3.6\pm0.32$ \citep{Mueller-06},
where $D$ is the distance of the galaxy in Mpc, $I$ is the \hii\
($1-0$) flux and $L$ is the \hii\ ($1-0$) luminosity. They are listed
in Table \ref{tab:lineratios}. Note that the latter conversion has
large uncertainties and that the ratio
$M_\mathrm{H_2}^\mathrm{hot}:M_\mathrm{H_2}^\mathrm{cold}$ spans at
least two orders of magnitude ($\sim10^{-7}$--$10^{-5}$). In addition,
$M_\mathrm{H_2}^\mathrm{hot}:M_\mathrm{H_2}^\mathrm{cold}$ depends on
the far-infrared colour $f_\nu(60\umu m)/f_\nu(100\umu m)$
\citep{Dale-05}, which may help to place tighter constraints on the
ratio. With a far-IR colour of $0.35$ \citep{Sakamoto-99} the ratio
would be approximately $10^{-8}$. This is consistent with the ratio of
$4\times10^{-7}$ obtained using the \citet{Mueller-06} conversion,
considering that both estimates have large uncertainties. A more
precise way to constrain the cold \hii\ masses is by more direct
measurements of CO ($J=1-0$) emission in the millimeter
range. \citet{Helfer-03} measured a peak molecular surface density
(i.e. the peak from the $6$~arcsec beam size) of
$815$~M$_\odot$~pc$^{-2}$. Over a $3$~arcsec aperture a total gas mass
of approximately $1.5\times10^7$~M$_\odot$ would then be
expected. \citet{Sakamoto-99} ($\sim3$~arcsec beam size) report a
molecular mass of $4\times10^{8}$~M$_\odot$
($2.67\times10^{8}$~M$_\odot$ when using the CO-to-\hii\ conversion
factor of \citealt{Helfer-03}) within a 15~arcsec diameter
aperture. Fig. \ref{fig:gasmass} shows the measured gas mass
distribution compared to (1) the dynamical mass distribution and (2)
the gas masses obtained from CO ($J=1-0$) measurements. We used figure
2 and equation 1 of \citet{Sakamoto-99} with the CO-to-\hii\
conversion factor of \citet{Helfer-03} to estimate the gas mass
distribution at smaller radii. Within a $3$~arcsec diameter aperture
the gas masses from \citet{Sakamoto-99} and \citet{Helfer-03} agree
very well. This shows that we overestimated the cold gas mass traced
by hot \hii\ by a factor of $\sim11$ when using the conversion of
\citet{Mueller-06}. When comparing the gas mass profile (calibrated to
match the gas mass at $r=1.5$~arcsec derived from CO) with the
dynamical mass, we find that within a $3$~arcsec aperture the gas is
approximately $5\%$ of the dynamical mass. At smaller radii this
fraction is even lower, as the gas mass profile is steeper than the
dynamical mass profile.  The gas mass distribution in quadrants $1$
and $4$ is larger by a factor $\sim1.5$ than in quadrants $2$ and $3$.
However, due to the small total mass we do not expect that the
irregular gas distribution has a large influence on the stellar
kinematics and it is unlikely the cause for the asymmetry in the
stellar kinematics; it might, however, be related to whatever is
responsible for the latter.

Concerning the results of the black hole mass measurement, the gas
mass has no significant influence. As the gas mass distribution has a
similar radial profile as the stellar mass, our
$\Upsilon_\mathrm{bulge}$ likely includes the gas mass, such that the
true stellar mass-to-light ratio is maybe slightly lower than our
nominal $\Upsilon_\mathrm{bulge}$. But this is not a serious problem
for the modelling. However, the fact that the gas is not evenly
distributed and the clouds already have a mass of order
$10^{6}$~M$_\odot$ each, can weaken the evidence for the presence of a
black hole, as it would imply that the centre is slightly out of
equilibrium.

\section{Dynamical Modelling of NGC\,3489}

\subsection{Construction of the Luminosity Profile for Modelling} 

Given the apparent similarity of NGC\,3489's inner structure to that
of NGC\,3368 (modulo the presence of a secondary bar in NGC\,3368),
including the strong isophotal twist created by the bar in NGC\,3489,
we followed a similar strategy for constructing the luminosity
profiles. That is, we divide the galaxy into separate disc (which
includes the discy pseudobulge) and central classical bulge
components, with the disc treated as having a constant observed
ellipticity of 0.41.  The disc surface brightness profile is an
azimuthal average with fixed ellipticity down to $r = 4.9$~arcsec,
with the profile at smaller radii being the extrapolated
inner-exponential fit from Fig.~\ref{fig:inner-decomp-n3489}.

The classical bulge profile is the result of a free-ellipse fit to the
inner-disc-subtracted WFPC2 F814W image. The latter was created by
generating a model disc with ellipticity = 0.41 and profile matching
the exponential part of the fit in Fig.~\ref{fig:inner-decomp-n3489}
(scale length = 4.9~arcsec), and then subtracting it from the
dust-corrected PC image.

The deprojection was done in the same way as for NGC\,3368 for bulge
and disc component separately.

\subsection{Dynamical Models}

NGC\,3489 has only a weak large-scale bar and no nuclear bar. The
measured kinematics and line indices are largely symmetric apart from
the asymmetry in $v$. Thus non-axisymmetries are not expected to play
a role as big as in NGC\,3368. We first use only SINFONI data to model
all four quadrants separately. However, we expect that, as for
NGC\,3368, due to the small field of view of the SINFONI data it will
be difficult to constrain $\Upsilon_{\mathrm{disc}}$, as the data
cover only that part of the galaxy where the classical bulge
dominates. Thus we try to constrain $\Upsilon_{\mathrm{disc}}$
beforehand by modelling SAURON and OASIS data alone. As the SAURON
data have a large field of view including the bar, we use just the
inner $10$~arcsec for that purpose. Finally, we model the combined
SINFONI plus OASIS and/or SAURON dataset.

We do not calculate models for different inclinations, as the
inclination is well determined from the photometry ($i=55\degr$). As
shown for NGC\,3368 in the previous section, the inclination cannot be
constrained better via dynamical modelling and the differences within
a small inclination range of a few degrees are small (see
Tab. \ref{tab:results} and Fig. \ref{fig:chi2_3368}).

\subsection{The stellar mass-to-light ratio of the disc}
\begin{figure}
\includegraphics[width=\linewidth,keepaspectratio]{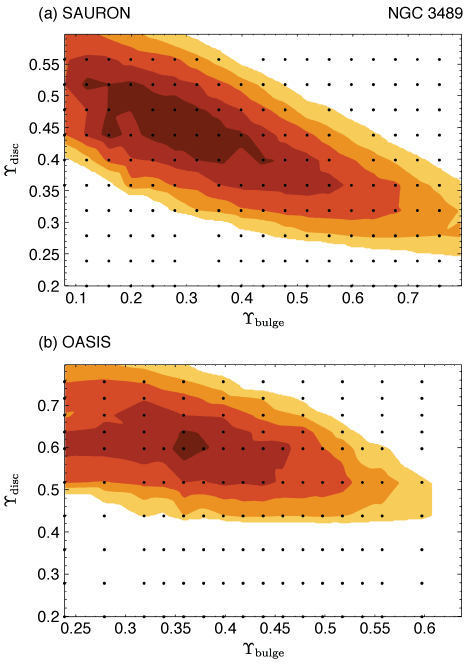}
\caption{$\Delta\chi_0^2=\chi^2-\chi_\mathrm{min}^2$ as a function of
$\Upsilon_\mathrm{disc}$ and $\Upsilon_{\mathrm{bulge}}$ for (a) the
symmetrised SAURON data and (b) the symmetrised OASIS data of
NGC\,3489. The coloured regions are the $1-5$~$\sigma$ confidence
intervals for two degrees of freedom. Each calculated model is marked
as a black dot. \label{fig:sauron}}
\end{figure}

In order to constrain $\Upsilon_\mathrm{disc}$ we first calculate
models using symmetrised SAURON and OASIS kinematics separately. We
only calculate models with $M_\bullet=0$ for the SAURON data, but vary
$M_\bullet$ between $0$ and $1.3\times10^7$~M$_\odot$ for the OASIS
data. Fig. \ref{fig:sauron}a shows $\Delta\chi^2$ as a function of
$\Upsilon_{\mathrm{bulge}}$ and $\Upsilon_{\mathrm{disc}}$ for the
SAURON models. $\Upsilon_{\mathrm{disc}}$ is well constrained, but in
$\Upsilon_{\mathrm{bulge}}$ a very large range between $\sim0$ and
$\sim0.68$ is possible. This is due to the fact that the classical
bulge is only just resolved with SAURON
($R_e^{\mathrm{CB}}=1.3$~arcsec, SAURON spatial resolution
$=1.1$~arcsec). The best-fitting model has
$\Upsilon_{\mathrm{bulge}}=0.28$ and
$\Upsilon_{\mathrm{disc}}=0.44$. For the OASIS models
(Fig. \ref{fig:sauron}b) the resulting $\Upsilon_{\mathrm{disc}}$ is
higher, which could be a result of the higher $\sigma$ of the OASIS
data compared to SAURON. Due to the higher spatial resolution
($0.69$~arcsec), $\Upsilon_{\mathrm{bulge}}$ is better
constrained. The best-fitting model has
$\Upsilon_{\mathrm{disc}}=0.6$ and
$\Upsilon_{\mathrm{bulge}}=0.36$. It is not possible to constrain
$M_\bullet$ with the OASIS data alone (see Fig. \ref{fig:chi2_3489}).

\subsection{The black hole mass}
To derive the mass of the SMBH we first use the SINFONI kinematics
alone to model the four quadrants separately.  We chose a few values
for $\Upsilon_{\mathrm{disc}}$ around $0.44$.  The results with the
corresponding $3\sigma$ errors are given in Table
\ref{tab:results3489}. The mean black hole mass for the four quadrants
is $\langle M_\bullet \rangle=4.25\times10^{6}$~M$_\odot$
(rms($M_\bullet$)$=2.05\times10^6$~M$_\odot$). $M_\bullet$ clearly
anticorrelates with $\Upsilon_{\mathrm{bulge}}$, but as in the case of
NGC\,3368 it does not depend on the specific choice of
$\Upsilon_{\mathrm{disc}}$. The mean black hole mass for any fixed
$\Upsilon_{\mathrm{disc}}$ is consistent with the result for any other
$\Upsilon_{\mathrm{disc}}$ within $1\sigma$.

\begin{figure*}
\includegraphics[width=1.0\linewidth,keepaspectratio]{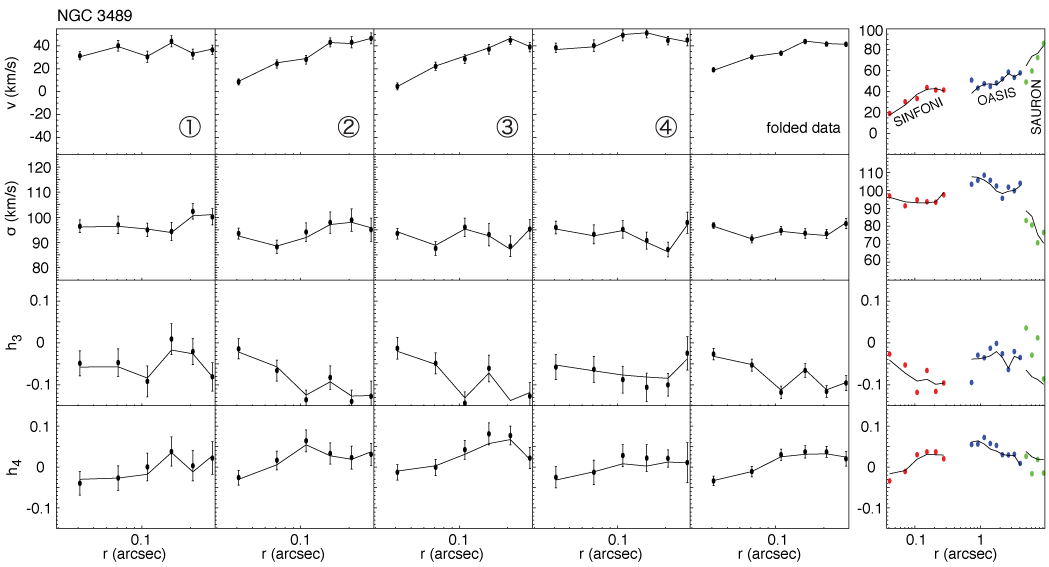}
\caption{The major axis kinematics of NGC\,3489 is shown as black
points for quadrants 1 to 4, the folded data and (in the last column)
the folded SINFONI, OASIS and SAURON data. Overplotted is the fit of
the best models with black hole.
\label{fig:fits3489}}
\end{figure*}

The error bars are large, such that
a wide range of black hole masses is allowed. A solution without
black hole is allowed in three quadrants within $2-4\sigma$ and in one
quadrant even within $1\sigma$. Thus there is no evidence for the
presence of a SMBH in one quadrant, and only weak evidence in the
others, when modelling the SINFONI data alone.
The fit of the best model in
each quadrant to $v$, $\sigma$, $h_3$ and $h_4$ along the major axis
is shown in Fig. \ref{fig:fits3489}.

The resulting $M_\bullet$, $\Upsilon_{\mathrm{bulge}}$ and
$\Upsilon_{\mathrm{disc}}$ of the four quadrants agree with each other
within $<2$~$\sigma$, and as there are also no obvious strong
inconsistencies between the kinematics of the quadrants, we fold the
LOSVDs of the four quadrants (the LOSVDs of quadrants 1 and 4 were
also flipped, such that $v$ and $h_3$ change sign). For the folded
data we find a best-fitting black hole mass of
$M_\bullet=5.0\times10^{6}$~M$_\odot$ at
$\Upsilon_{\mathrm{bulge}}=0.56$. This is in good agreement with the
results of the individual quadrants. A solution without black hole is
allowed within $3\sigma$, thus as a conservative result we can only
give a $3\sigma$ upper limit of $1.3\times10^{7}$~M$_\odot$ for the
SMBH in NGC\,3489, when using just the SINFONI data. 

The non-dependence of $M_\bullet$ on $\Upsilon_{\mathrm{disc}}$ can be
explained by the very small field of view of the SINFONI data, which
covers only the very central part of the galaxy, dominated  by the
classical bulge. This might also explain the relatively weak detection
of a SMBH in NGC\,3489 despite the high quality data. It therefore
seems  reasonable to  include  kinematics at  larger  radii, like  the
SAURON or the OASIS kinematics, as these datasets cover a large
fraction of the disc and therefore are able to constrain
$\Upsilon_{\mathrm{disc}}$ very well, as shown above. We should keep
in mind however, that the SAURON and OASIS velocity dispersions do not
fully agree with each other, are larger than the  SINFONI dispersion
and show some deviations from axisymmetry, which might possibly be due
to the  strong dust features. In  order to determine  how strong these
differences affect  the result  of the modelling  we do three  sets of
models: the  first one with SINFONI  and OASIS data  (using OASIS data
between $r=0.5$~arcsec and 4~arcsec),  the second one with SINFONI and
SAURON data (using SAURON data between $r=1$~arcsec and 10~arcsec) and
the  third  one with  all  three  datasets  (with OASIS  data  between
$r=0.5$~arcsec  and  4~arcsec and  SAURON  data  between 4~arcsec  and
10~arcsec).

\begin{figure}
\includegraphics[width=\linewidth,keepaspectratio]{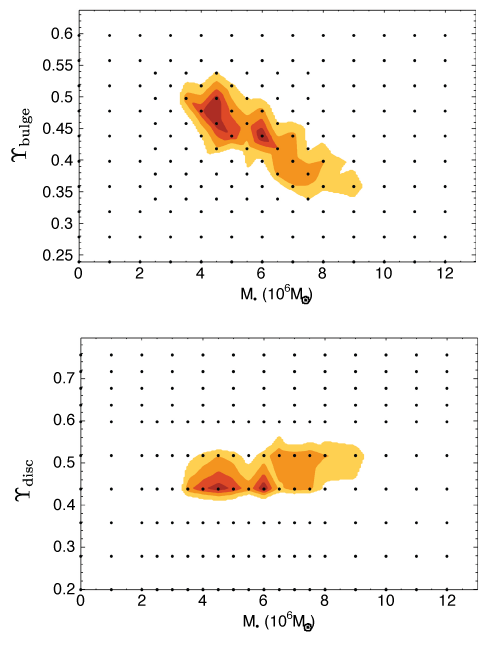}
\caption{Same as Fig. \ref{fig:models} for NGC\,3489, with an
inclination $i=55\degr$. The averaged SINFONI data, OASIS data between
$0.5$ and $4$~arcsec and SAURON data between $4$ and $10$~arcsec were
used for the modelling. \label{fig:models3489}}
\end{figure}

\begin{figure*}
\includegraphics[width=0.8\linewidth,keepaspectratio]{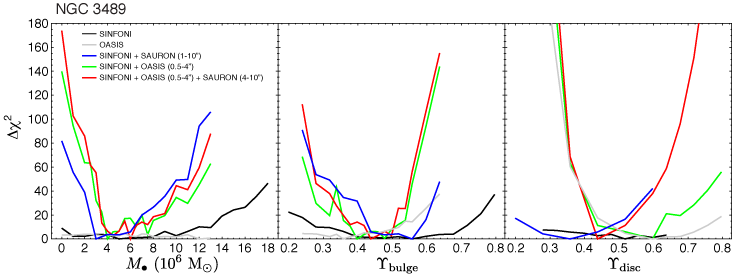}
\caption{$\Delta\chi_0^2=\chi^2-\chi_{\mathrm{min}}^2$ as a function
of $M_\bullet$ (left column, marginalised over
$\Upsilon_{\mathrm{bulge}}$ and $\Upsilon_{\mathrm{disc}}$),
$\Upsilon_{\mathrm{bulge}}$ (middle column, marginalised over
$M_\bullet$ and $\Upsilon_{\mathrm{disc}}$) and
$\Upsilon_{\mathrm{disc}}$ (right column, marginalised over
$M_\bullet$ and $\Upsilon_{\mathrm{bulge}}$) for NGC\,3489. The
colours indicate the datasets used for the modelling (black: SINFONI,
grey: OASIS, blue: SINFONI+SAURON, green: SINFONI+OASIS, red:
SINFONI+OASIS+SAURON).\label{fig:chi2_3489}}
\end{figure*}

Fig. \ref{fig:models3489} shows the resulting $\Delta\chi^2$
contours for the combined SINFONI, OASIS and SAURON data. The error
contours are very narrow and both $\Upsilon_\mathrm{bulge}$ and
$\Upsilon_\mathrm{disc}$ are very well constrained.
Fig. \ref{fig:chi2_3489} shows the resulting $\Delta\chi^2$ profiles
for all data combinations we used. It is clear that the mass of the
black hole can be much better constrained when including SAURON and/or
OASIS data. The constraints on $\Upsilon_{\mathrm{bulge}}$ and
$\Upsilon_{\mathrm{disc}}$ are also much stronger in these
cases. Using SAURON data in addition to SINFONI and OASIS does not
seem to improve the measurement of $M_\bullet$ and
$\Upsilon_{\mathrm{bulge}}$. The scatter in the $\Delta\chi^2$
profiles is quite large for the models of the combined datasets,
despite the good quality and high S/N of the individual datasets and
despite the comparatively small scatter in the models of individual
datasets. The uncertainties of the SMBH mass measurement therefore do
not seem to be dominated by statistical errors, but instead by
systematics. Systematic errors can be introduced e.g. due to the
differences in the kinematics of the individual datasets. Systematic
errors in the modelling (e.g. slightly different results for different
quadrants) could add to the scatter as well, but are difficult to
quantify. We measure the formal $1$~$\sigma$ errors (corresponding to
$\Delta\chi^2=1$ for one degree of freedom) by fitting a third order
polynomial to each curve in Fig. \ref{fig:chi2_3489}. The best values
for $M_\bullet$, $\Upsilon_{\mathrm{bulge}}$ and
$\Upsilon_{\mathrm{disc}}$ given in Table \ref{tab:results3489_q5}
refer to the minimum of the fit and the associated $\Delta\chi^2\leq1$
region. We then determine the systematic error, introduced by the
differences between the datasets, from the scatter of the best fits
for models with combined datasets. Thus when using all available data,
we obtain a black hole mass of
$M_\bullet=(6.00^{+0.56}_{-0.54}|_{\mathrm{stat}}\pm0.64|_{\mathrm{sys}})\times10^{6}$~M$_\odot$,
a bulge mass-to-light ratio
$\Upsilon_{\mathrm{bulge}}=0.45\pm0.02|_{\mathrm{stat}}\pm0.03|_{\mathrm{sys}}$
and a disc mass-to-light ratio
$\Upsilon_{\mathrm{disc}}=0.47^{+0.01}_{-0.02}|_{\mathrm{stat}}\pm0.05|_{\mathrm{sys}}$.

\begin{figure*}
\includegraphics[width=\linewidth,keepaspectratio]{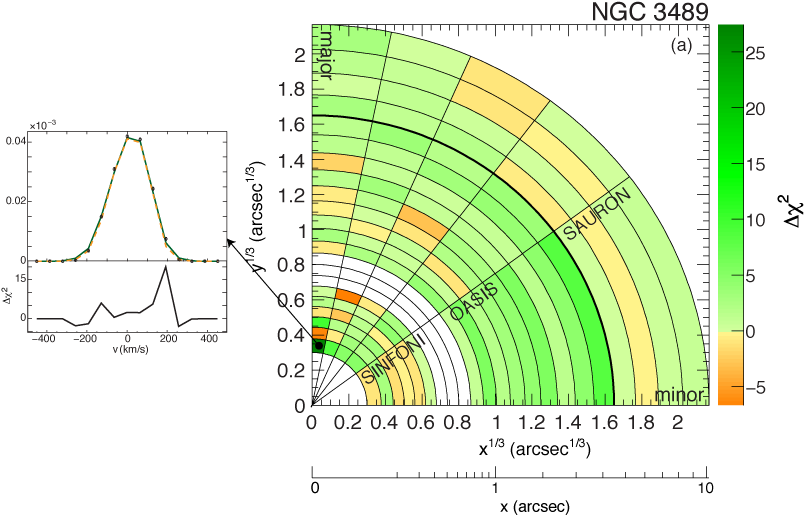}
\caption{Same as Fig. \ref{fig:green3368} for NGC\,3489 for the models using the averaged SINFONI+OASIS+SAURON data. Instead of a linear radial scale we used $x^{1/3}$ and $y^{1/3}$ due to the increasing bin size with radius. The conversion is shown in the bottom scale. \label{fig:green3489}}
\end{figure*}

\subsection{Evidence for a black hole in NGC\,3489}


In order to illustrate where the influence of the black hole is
largest, Fig. \ref{fig:green3489} shows the $\chi^2$ difference
between the best-fitting model without a black hole and the
best-fitting model with a black hole for all LOSVDs of the combined
SINFONI+OASIS+SAURON dataset.  The fit with black hole is generally
better in 80\% of all bins. Along the major axis the largest $\chi^2$
differences appear in the LOSVD wings, both at negative and positive
velocities. Improvements of the fit appear at all radii.  

\begin{figure}
\includegraphics[width=0.95\linewidth,keepaspectratio]{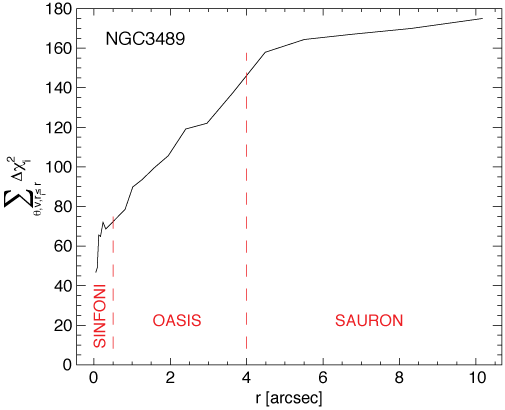}
\caption{Same as Fig. \ref{fig:radchi2_3368} for the
combined SINFONI+OASIS+SAURON data of NGC\,3489. \label{fig:radchi2_3489}}
\end{figure}

Fig. \ref{fig:radchi2_3489} shows $\Delta\chi^2$ summed over all
angles and velocities as a function of radius. The $\Delta\chi^2$
increase is steepest in the region covered by the SINFONI data
($\Delta\chi^2_\mathrm{SINFONI}\approx72$, mostly coming from the
region $\lesssim0.15$~arcsec, corresponding to $\sim3r_\mathrm{SoI}$).
It then grows by about the same amount in the region of the OASIS data
($0.5-4$~arcsec). At larger radii (region of the SAURON data) the
$\Delta\chi^2$ increase is only small.

Although with the OASIS data alone it is not possible to constrain the
mass of the SMBH, the region covered by this dataset
($\sim0.5-4$~arcsec) seems to be crucial for the lower limit on
$M_\bullet$, which is not possible to derive with the SINFONI data
alone. This means that differences between models without black hole and
models with black hole (say, $M_\bullet=6\times10^6$~M$_\odot$) should
not only appear within the sphere of influence, but also further
outside. This should not be surprising.  For example, some of the
effects of a central mass concentration in an isotropic system can be
mimicked in a system without such a concentration by enhanced radial
anisotropy \citep[e.g.,][]{Binney-82}. In such a case, the region
where radial anistropy is required extends well outside the nominal
sphere of influence of the central mass.  Only in cases where the
best-fit model without a black hole has exactly the same orbital
structure as the best-fit model with the black hole would differences
between the fits be (roughly) confined to the sphere of influence. 
This is in agreement with the observations in NGC\,3368, Fornax\,A and
NGC\,4486a \citep{Nowak-07,Nowak-08}, where a general improvement of the
fit even outside the sphere of influence, was observed.

The total $\Delta\chi^2$, summed over all
LOSVDs, between the best-fitting model without a black hole and the
best-fitting model with a black hole is given in the last column of
Table \ref{tab:results3489_q5}.

\subsection{Discussion}

No attempts have been made in the literature to measure the mass of the
SMBH in NGC\,3489. From the $M_\bullet$-$\sigma$ relation of
\citet{Tremaine-02} we would expect a mass between
$M_{\bullet}=5.2\times10^{6}$~M$_\odot$ for
$\sigma_\mathrm{e}=88.9$~\kms\ derived from the SAURON data
\citep{Emsellem-04} and $M_\bullet=9.2\times10^6$~M$_\odot$ for
$\sigma_\mathrm{e}=102.5$~\kms\ derived from the OASIS data
\citep{McDermid-06}.  From the relation between $M_\bullet$ and $K$-band
magnitude \citep{MarconiHunt-03} we would expect a black-hole mass of
$M_{\bullet}=1.92\times10^7$~M$_\odot$ if it correlates with the total
(photometric) bulge magnitude $M_{K,\mathrm{total}}=-21.91$, or
$4.94\times10^6$~M$_\odot$ if it correlates with the classical bulge
magnitude $M_{K,\mathrm{bulge}}=-20.60$ only.

The stellar population models of \citet{Maraston-98,Maraston-05}
suggest an age of $\sim1$~Gyr for the best-fitting
$\Upsilon_\mathrm{bulge}=0.45\approx\Upsilon_\mathrm{disc}$ and a high
metallicity [Z/H]$\sim0.67$, and an age of $\sim2-3$~Gyr for a solar
metallicity population (assuming a Salpeter IMF). This is in agreement
with \citet{McDermid-06}, who find both an age gradient (from
$\sim2-3$~Gyr in the outer regions to $\sim1$~Gyr in the centre) and a
metallicity gradient (from $\sim$solar in the outer regions to
$\sim0.6$ in the centre). It is also compatible with \citet{Sarzi-05},
who found a mean age of $\sim3$~Gyr assuming solar metallicity. The
stellar mass within $r_\mathrm{SoI}\approx0.053$~arcsec is
$\approx6\times10^6$~M$_\odot$. If the best-fitting mass for the black
hole of $M_{\bullet}=6.0\times10^6$~M$_{\odot}$ were entirely composed
of stars, the mass-to-light ratio would increase to $0.9$.  This would
be typical for an older stellar population ($\sim6$~Gyr for a high
metallicity [Z/H]=0.67 and a Salpeter IMF), and therefore conflict
with the values found by \citet{Sarzi-05} and \citet{McDermid-06}.

\begin{table*}
 \centering
 \begin{minipage}{140mm}
  \caption{Resulting black hole masses $M_\bullet$ and $H$-band mass-to-light ratios $\Upsilon_{\mathrm{bulge}}$ and $\Upsilon_{\mathrm{disc}}$ of NGC\,3489. The lower and upper $3\sigma$ limits are given in brackets. The total $\chi^2$ of the best model with black hole and the $\chi^2$ difference between the best model without black hole and the best model with black hole are given in the last two columns.}\label{tab:results3489}
  \begin{tabular}{llllll}
  \hline
Quadrant & $M_\bullet$ [$10^6$~M$_\odot$]  & $\Upsilon_\mathrm{bulge}$ & $\Upsilon_{\mathrm{disc}}$  & $\chi_{\mathrm{min}}^{2}$ & $\Delta\chi^2_\mathrm{noBH-BH}$ \\
 \hline
 1        &  4.0 (0.0, 8.0)  &  0.60 (0.44, 0.76) & 0.36 (0.28, 0.64)  & 106.812 & 3.289\\
 2        &  1.0 (0.0, 7.0)  &  0.60 (0.44, 0.80) & 0.60 (0.28, 0.64)  & 100.554 & 0.301 \\
 3        &  6.0 (0.0, 13.0) &  0.48 (0.28, 0.72) & 0.52 (0.28, 0.64)  & 57.612  & 10.487\\
 4        &  6.0 (1.0, 10.0) &  0.52 (0.36, 0.64) & 0.32 (0.28, 0.64)  & 80.077  & 16.786\\
 folded        &  5.0 (0.0, 13.0) &  0.56 (0.28, 0.72) & 0.52 (0.28, 0.64)  & 47.877  & 9.060\\
\hline
\end{tabular}
\end{minipage}
\end{table*}

\begin{table*}
 \centering
 \begin{minipage}{140mm}
  \caption{Resulting black hole masses $M_\bullet$ and $H$-band
  mass-to-light ratios $\Upsilon_{\mathrm{bulge}}$ and
  $\Upsilon_{\mathrm{disc}}$ of NGC\,3489 for the folded SINFONI data
  alone and in combination with SAURON and OASIS kinematics. The lower
  and upper $1\sigma$ limits (1 degree of freedom), determined by
  fitting a third order polynomial to the $\Delta\chi^2$ profiles of
  Fig. \ref{fig:chi2_3489}, are given in brackets. The total $\chi^2$
  of the best model with black hole and the $\chi^2$ difference
  between the best model without black hole and the best model with
  black hole are given in the last two
  columns.}\label{tab:results3489_q5}
  \begin{tabular}{llllll}
  \hline
 & $M_\bullet$ [$10^6$~M$_\odot$]  & $\Upsilon_\mathrm{bulge}$ & $\Upsilon_{\mathrm{disc}}$  & $\chi_{\mathrm{min}}^{2}$ & $\Delta\chi^2$ \\
 \hline
 SINFONI                                                  &  5.97 (3.64, 8.13) & 0.53 (0.48, 0.58) & 0.55 (0.48, 0.62) & 106.812 & 3.289 \\
 SINFONI + SAURON($1-10\arcsec$)                          &  4.56 (4.03, 5.12) & 0.52 (0.50, 0.54) & 0.36 (0.32, 0.40) & 368.458 & 81.83\\
 SINFONI + OASIS($0.5-4\arcsec$)                          &  5.81 (5.21, 6.46) & 0.46 (0.44, 0.48) & 0.52 (0.50, 0.55) & 551.297 & 139.86\\
 SINFONI + OASIS($0.5-4\arcsec$) + SAURON ($4-10\arcsec$) &  6.00 (5.46, 6.56) & 0.45 (0.43, 0.47) & 0.47 (0.45, 0.48) & 606.653 & 173.86\\
\hline
\end{tabular}
\end{minipage}
\end{table*}

\section{Summary and Discussion}

We analysed near-IR integral-field data for two barred galaxies that
host both a pseudobulge and a classical bulge component. Both galaxies
show fast and regular rotation and a $\sigma$-drop at the centre,
which in the case of NGC\,3368 is more pronounced and may have
developed from gas, transported to the inner region by the bars and
spiral arms.  The kinematics of NGC\,3368 -- in particular the
velocity dispersion -- is asymmetric.  The reasons for that could be
dust or (less likely) the non-axisymmetric potential induced by the
two bars. The gas distribution is also inhomogeneous, but as the total
gas mass accounts for only $\la5\%$ of the dynamical mass, this has
probably no significant influence on the stellar kinematics.  There
are two kinematically decoupled gas clouds located a few tens of
parsecs north of the centre. Each cloud has a total mass of order
$10^6$M$_\odot$.  The stellar kinematics of NGC\,3489 is very regular,
with a slight asymmetry in the velocity field.  All other kinematic
parameters and the line indices are consistent with axisymmetry. No
gas emission was found in NGC\,3489. The near-IR line indices \nai\
and CO show a negative gradient in both galaxies, indicating an age
and/or metallicity gradient.

We applied axisymmetric dynamical models to derive the SMBH masses in
NGC\,3368 and NGC\,3489. In our models we assume that the galaxy
potential can be decomposed into three components: the central black
hole, an inner, classical bulge (with mass-to-light ratio
$\Upsilon_\mathrm{bulge}$) and the disc ($\Upsilon_\mathrm{disc}$);
the disc component includes the pseudobulge. The inclination of the
models is fixed by the isophotes of the outer disc. For NGC\,3368 we
modelled the four quadrants of our IFU data independently and the
resulting black hole masses and mass-to-light ratios agree very
well. We find that $M_\bullet$ is largely independent of
$\Upsilon_{\mathrm{disc}}$ and anticorrelates with
$\Upsilon_{\mathrm{bulge}}$. The average black hole mass for the four
quadrants and an inclination $i=53\degr$ is $\langle
M_\bullet\rangle=7.5\times10^{6}$~M$_\odot$
(rms($M_\bullet$)$=1.5\times10^{6}$~M$_\odot$). A solution without a
black hole is excluded by $\approx4$--$5\sigma$. The errors,
however, cover a large range in $M_\bullet$.  The largest uncertainty
for $M_\bullet$ comes from the unknown $\Upsilon_{\mathrm{bulge}}$,
and independent constraints, e.g. from stellar population modelling,
would likely improve the results. However, unless the shape of the IMF
is known, mass-to-light ratios from stellar population analyses are
ambiguous. The scatter from quadrant to quadrant is smaller than the
uncertainty related to $\Upsilon_\mathrm{bulge}$, suggesting that the
symmetry assumption plays a minor role for the uncertainty of
$M_\bullet$.  Our results do not significantly depend on the
inclination (within the photometrically allowed inclination range).

For NGC\,3489, modelling of the four SINFONI quadrants likewise gave
consistent black hole masses and mass-to-light ratios. Similar to
NGC\,3368 the errors in $M_\bullet$ are large, the
black hole mass is independent of $\Upsilon_{\mathrm{disc}}$ and it
clearly anticorrelates with $\Upsilon_{\mathrm{bulge}}$. Modelling the
folded SINFONI data gives the same result as for the individual
quadrants; thus, non-axisymmetries do not seem to play a role. When
including OASIS and/or SAURON data, $\Upsilon_\mathrm{bulge}$ and
therefore also $M_\bullet$ could be much better constrained.  Using
all three datasets, we derived for NGC\,3489 a SMBH mass of
$M_\bullet=(6.00^{+0.56}_{-0.54}|_{\mathrm{stat}}\pm0.64|_{\mathrm{sys}})\times10^{6}$~M$_\odot$
with a bulge mass-to-light ratio of
$\Upsilon_{\mathrm{bulge}}=0.45\pm0.02|_{\mathrm{stat}}\pm0.03|_{\mathrm{sys}}$
and a disc mass-to-light ratio
$\Upsilon_{\mathrm{disc}}=0.47^{+0.01}_{-0.02}|_{\mathrm{stat}}\pm0.05|_{\mathrm{sys}}$. A
solution without a black hole is excluded with high significance. To
derive a firm lower limit to $M_\bullet$, data between
$\sim0.5-4$~arcsec seem to be crucial, in addition to the
high-resolution SINFONI data in the centre. With OASIS data alone, no
limits on $M_\bullet$ could be placed. There are some inconsistencies
in the kinematics between the three datasets, which seem to be the
main source of systematic errors. In particular when modelling OASIS
data alone, we get a higher $\Upsilon_\mathrm{disc}$ than if modelling SAURON
data alone (because the inner $\sigma$ is higher in the OASIS data
than in the SAURON data).

\begin{figure*}
\includegraphics[width=\linewidth,keepaspectratio]{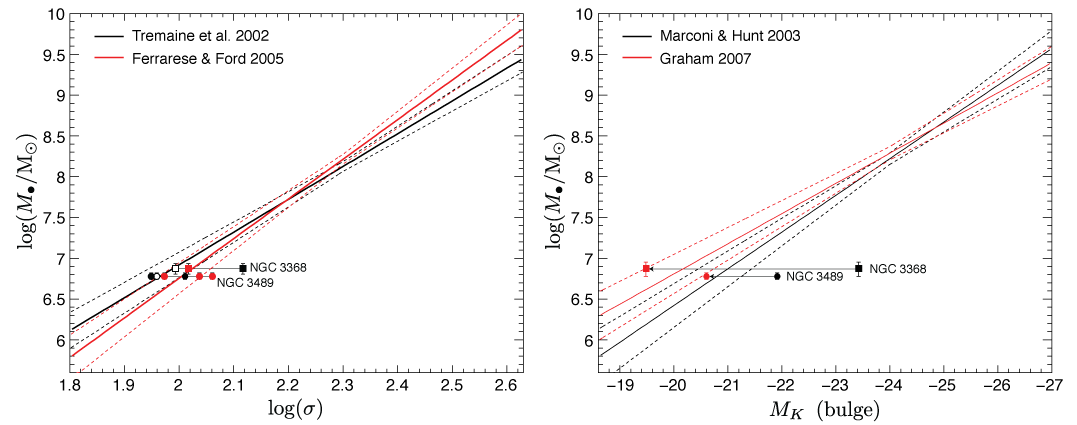}
\caption{Left panel: Location of NGC\,3368 and NGC\,3489 with respect
to the $M_{\bullet}$-$\sigma$ relation (black: \citealt{Tremaine-02},
red: \citealt{Ferrarese-05}). The velocity dispersion from our SINFONI
measurements (open symbols) and values for $\sigma_{\mathrm{e}}$
(filled black symbols) and $\sigma_\mathrm{e/8}$ (filled red symbols)
derived from the literature are plotted for each galaxy. Right panel:
Location of the two galaxies with respect to the $M_\bullet$-$M_K$
relation of \citet{MarconiHunt-03} (black) and \citet{Graham-07b}
(red). The $K$-band magnitudes of the total photometric bulges are
plotted as filled black symbols, the magnitudes of the classical bulge
components as filled red symbols. In both panels the average
$M_\bullet$ of the four quadrants and the rms is plotted for
NGC\,3368, and $M_\bullet$ from the combination of SINFONI, SAURON and
OASIS data with its statistical $1\sigma$ error is plotted for
NGC\,3489. \label{fig:msigma}}
\end{figure*}

The implications for the $M_\bullet$-$\sigma$ relation and the
$M_\bullet$-$M_K$ relation are illustrated in
Fig. \ref{fig:msigma}. For NGC\,3368 the mean $M_\bullet$ of the four
quadrants and the rms, and for NGC\,3489 $M_\bullet$ from the
combination of SINFONI, SAURON and OASIS data with its statistical
$1\sigma$ error is plotted against $\sigma$ and $M_K$ using the
relations of \citet{Tremaine-02}, \citet{Ferrarese-05},
\citet{MarconiHunt-03} and \citet{Graham-07b}. All values for
$\sigma_{\mathrm{e}}$ and $\sigma_{\mathrm{e/8}}$ were measured using
the effective radius of the total photometric bulge, as was done for
all the galaxies contributing to the \citet{Tremaine-02} and
\citet{Ferrarese-05} relations. No attempt to determine
$\sigma_\mathrm{e}$ for the classical components has therefore been
made, but as we use luminosity-weighted measurements, all values
determined from high-resolution data represent mostly the classical
bulge.

The agreement of NGC\,3368 with the $M_\bullet$-$\sigma$ relation
largely depends on the value of $\sigma$ which is used. The small
$\sigma=98.5$~\kms\ measured within the SINFONI field of view is in
good agreement with the $M_\bullet$-$\sigma$ relation. When combining
the SINFONI $\sigma$ with $\sigma$ measurements of
\citet*{Whitmore-79}, \citet{Heraudeau-99} and \citet{VegaBeltran-01}
a value of $\sigma_\mathrm{e/8}=104$~\kms\ is obtained. The velocity
dispersions from the literature alone however
(e.g. $\sigma_{\mathrm{e}}=117$~\kms\ estimated by \citealt{Sarzi-02},
$\sigma_{\mathrm{e}}=130.9$~\kms\ respectively
$\sigma_{\mathrm{e/8}}=129.9$~\kms\ measured by
\citealt{Heraudeau-99}, or $\sigma\approx150$~\kms\ by
\citealt{Moiseev-04}) are significantly larger than expected by this
estimate and not or only marginally in agreement with the
$M_\bullet$-$\sigma$ relation. With a $K$-band magnitude of $-23.42$
for the total photometric bulge, NGC\,3368 falls far (a factor of
$\sim12$) below the $M_\bullet$-$M_K$ relation of
\citet{MarconiHunt-03}. If we postulate that the SMBH only correlates
with the magnitude of the classical bulge, the situation
improves. With $M_K^{\mathrm{CB}} = -19.48$ NGC\,3368 now lies a
factor of $\sim5$ above the $M_\bullet$-$M_K$ relation of
\citet{MarconiHunt-03}, but is in good agreement with the
$M_\bullet$-$M_K$ relation of \citet{Graham-07b}.

For NGC\,3489 the situation is similar. $M_\bullet$ is in excellent
agreement with the $M_\bullet$-$\sigma$ relation when using either the
SINFONI mean $\sigma=91.1$~\kms\ or the SAURON values
$\sigma_\mathrm{e}=88.9$~\kms\ and $\sigma_\mathrm{e/8}=94$~\kms. It
is still in reasonably good agreement with the relation when using the
OASIS measurements ($\sigma_\mathrm{e}=102.5$~\kms,
$\sigma_\mathrm{e/8}=108.9$~\kms) or when taking into account other
$\sigma$ measurements from the literature
($\sigma_\mathrm{e/8}=115$~\kms\ using
\citealt{Whitmore-79,DalleOre-91,Smith-00,Barth-02} and the SINFONI
value).  With a $K$-band magnitude of the total photometric bulge of
$M_K^{\mathrm{PB}}=-21.91$ NGC\,3489 also falls far below the
$M_\bullet$-$M_K$ relation of \citet{MarconiHunt-03} and
\citet{Graham-07b}, but is in excellent agreement if the magnitude of
the classical bulge component is considered ($M_K=-20.60$).

The large difference in the $\sigma$ measurements makes it difficult
to draw any firm conclusion with respect to the location of
pseudobulges in the $M_\bullet$-$\sigma$ relation, and at the same
time illustrates that measurement errors in $\sigma$ may play a larger
role than one may have thought, in particular when dealing with small
galaxy samples. NGC\,3368 would fall far below the
$M_\bullet$-$\sigma$ relation when optical longslit kinematics alone
are used. These discrepancies between the $\sigma$ measurements might
at least partly be due to dust, which affects the optical data much
more than the near-IR data.

The $K$-band magnitudes on the other hand can be determined very
accurately even for subcomponents of the galaxy. Taken at face value,
both galaxies clearly do not follow the $M_\bullet$-$M_K$ relation of
\citet{MarconiHunt-03} when considering the $K$-band magnitudes of the
total photometric bulge, but are in better (NGC\,3368) or even
excellent (NGC\,3489) agreement with it when considering the classical
bulge magnitude only. 

If we take into account that a stellar population becomes fainter when
it ages passively ($2.3$~mag in $K$ band for a solar metallicity
population between $1$ and $10$~Gyr, based on the stellar population
models of \citealt{Maraston-98,Maraston-05}), the pseudobulges would
move toward the $M_\bullet$-$M_K$ relation with time. Given the
uncertainties on the age estimate, the exact size of the effect is
unclear. Keeping in mind this caveat, this is in line with
\citet{Greene-08}, who conclude that pseudobulges follow the
$M_\bullet$-$\sigma$ relation, but not the
$M_\bullet$-$M_\mathrm{bulge}$ relation, as well as with
\citet{Gadotti-08}, who find that pseudobulges follow only one of the
two relations, if any. In order to strengthen our results, studies of
a larger sample of pseudobulges similar in design are necessary.

Whether modelling single quadrants of obviously non-axisymmetric
galaxies with an axisymmetric code is a good approximation and gives
the correct black hole masses is certainly still an issue that remains
to be resolved. The recently developed triaxial codes of
\citet{deLorenzi-07} and \citet{vandenBosch-08} will have
the potential to solve this issue in the future.

\section*{Acknowledgments}
We would like to thank the Paranal Observatory Team for support during
the observations. We are grateful to Harald Kuntschner and Mariya
Lyubenova for providing us the code to measure near-IR line indices,
and to Karl Gebhardt for providing the MPL code. Furthermore we thank
Alexei Moiseev and Richard McDermid for providing us their 2D
kinematics on NGC\,3368 and NGC\,3489. We would also like to
thank Maximilian Fabricius, Roland Jesseit and Erin Hicks for valuable
discussions. Finally we would like to thank the referee Eric Emsellem
for his critical comments which helped us to improve the
manuscript. This work was supported by the Cluster of Excellence:
``Origin and Structure of the Universe'' and by the Priority Programme
1177 ``Galaxy Evolution'' of the Deutsche Forschungsgemeinschaft.

\bibliographystyle{mn2e}
\bibliography{bibliography}

\appendix


\bsp

\label{lastpage}

\end{document}